\documentclass[traditabstract]{aa}
\usepackage{epsfig}    
\usepackage{lscape}
\usepackage{natbib}
\bibpunct{(}{)}{;}{a}{}{,}    
\usepackage{graphics}
\usepackage{graphicx,graphics}
\usepackage{color} 
\usepackage{times}
\usepackage{amsmath}
\usepackage{amssymb}
\begin{document}

\title{Breaks in surface brightness profiles and radial abundance gradients in the discs of spiral galaxies} 
      
\author {
        L.~S.~Pilyugin\inst{\ref{MAO},\ref{ARI}} \and 
        E.~K.~Grebel\inst{\ref{ARI}} \and
        I.~A.~Zinchenko\inst{\ref{MAO},\ref{ARI}} \and
        Y.~A.~Nefedyev\inst{\ref{KGU}} \and
        J.~M.~V\'{i}lchez\inst{\ref{IAA}}
\institute{Main Astronomical Observatory, National Academy of Sciences of Ukraine, 27 Akademika Zabolotnoho St, 03680, Kiev, Ukraine \label{MAO} \and
Astronomisches Rechen-Institut, Zentrum f\"{u}r Astronomie der Universit\"{a}t Heidelberg, M\"{o}nchhofstr.\ 12--14, 69120 Heidelberg, Germany \label{ARI} \and
Kazan Federal University, 18 Kremlyovskaya St., 420008, Kazan, Russian Federation \label{KGU} \and
Instituto de Astrof\'{\i}sica de Andaluc\'{\i}a, CSIC, Apdo 3004, 18080 Granada, Spain \label{IAA} }}

\abstract{
We examine the relation between breaks in the surface brightness
profiles and radial abundance gradients within the optical radius  
in the discs of 134
spiral galaxies from the CALIFA survey.  The distribution of the
radial abundance (in logarithmic scale) in each galaxy was fitted by
simple and broken linear relations. The surface brightness profile was
fitted assuming pure and broken exponents for the disc.  We find that
the maximum absolute difference between the abundances in a disc given
by broken and pure linear relations is less than 0.05 dex in the
majority of our galaxies and exceeds the scatter in abundances
for 26 out of 134 galaxies considered.    
The scatter in abundances around the broken
linear relation is close (within a few percent) to that around the
pure linear relation.  The breaks in the surface brightness profiles
are more prominent.  The scatter around the broken exponent in a
number of galaxies is lower by a factor of two or more than that
around the pure exponent.  The shapes of the abundance gradients and
surface brightness profiles within the optical radius 
in a galaxy may be different. A pure
exponential surface brightness profile may be accompanied by a broken
abundance gradient and vise versa. There is no correlation between
the break radii of the abundance gradients and surface brightness
profiles. Thus, a break in the surface brightness profile does not
need to be accompanied by a break in the abundance gradient.
}

\keywords{galaxies: abundances -- ISM: abundances -- H\,{\sc ii} regions, galaxies}

\titlerunning{Breaks in disc galaxy abundance gradients}
\authorrunning{Pilyugin et al.}
\maketitle

\section{Introduction}

It was established many years ago
\citep{Vaucouleurs1959,Freeman1970,vanderKruit1979} that the surface
brightness profiles of the discs of spiral galaxies can be fitted by
an exponential.  However, pure exponential functions can give only
a rough representation of the radial surface brightness profiles of
the discs of some galaxies. \citet{Pohlen2006} found that only around 10
to 15 per cent of all spiral galaxies have a normal/standard purely
exponential disc, while the surface brightness distribution of the rest
of the galaxies is better described by a broken exponential.  

It is also well-known that discs of spiral galaxies show negative
radial abundance gradients in the sense that the abundance is higher
at the centre and decreases with galactocentric distance
\citep{Searle1971,Smith1975}.  It is common practice to express the
oxygen abundance in a galaxy using the logarithmic scale, which
permits one to describe the abundance distribution across the disc of
a galaxy with a linear relation between oxygen abundance O/H and
galactocentric distance $R{_g}$   
; we used fractional galactocentric distances $R{_g}$ normalized to the 
optical isophotal radius $R_{25}$ in the current study.  

The variation in the (logarithm of)
abundance with radius has been examined for many galaxies by different
authors \citep[][among many
others]{VilaCostas1992,Zaritsky1994,Ryder1995,vanZee1998,
Pilyugin2004,Pilyugin2006,Pilyugin2007,Pilyugin2014a,Pilyugin2015,
Moustakas2010,Gusev2012,Sanchez2014,Zinchenko2015,Ho2015,BresolinKennicutt2015}.
Previous studies found that the gradients are reasonably well fitted
by a straight line although in some spiral galaxies the slope appears
to steepen (or flatten) in the central regions \citep[][among many
others]{VilaCostas1992,Zaritsky1992,MartinRoy1995,Zahid2011,Scarano2011,Sanchez2014,Zinchenko2016}.
For example, the abundance gradient in the disc of the Milky Way
traced by Cepheid abundances seems to flatten in the central region
\citep{Martin2015,Andrievsky2016}.

There is a correlation between the local oxygen abundance and stellar surface brightness in the discs of spiral galaxies
\citep[e.g.][]{Webster1983,Edmunds1984,Ryder1995,Moran2012,RosalesOrtegaetal2012,Pilyugin2014b}.
If this relation is indeed local, i.e. if there is a point-to-point
relation between the surface brightness and abundance, then one can
expect that the break in the surface brightness distribution should be
accompanied by a break in the abundance distribution.   

The existence of breaks in
gradients reported in earlier works can be questioned for two reasons.
Firstly, those gradients are typically based on a small number of
H\,{\sc ii} region abundances.  Secondly, it has been recognized that
the shape of the abundance distribution is very sensitive to the
calibration used for the abundance determinations
\citep[e.g.][]{Zaritsky1994,KennicuttGarnett1996,Pilyugin2001,Pilyugin2003}.
 \citet{Zaritsky1994} pointed out that the question of the shape of
the abundance gradient remains open and can be solved only when
measurements of many  H\,{\sc ii} region abundances per galaxy and a
reliable method for abundance determinations become available.  

Recently, measurements of spectra of dozens of H\,{\sc ii} regions in
several galaxies were obtained
\citep[e.g.][]{Berg2015,Croxall2015,Croxall2016}.  In the framework
of surveys such as the Calar Alto Legacy Integral Field Area
(CALIFA) survey \citep{Sanchez2012, Husemann2013, GarciaBenito2015} and
the Mapping Nearby Galaxies at Apache Point Observatory (MaNGA)
survey \citep{Bundy2015}, 2D spectroscopy of a large
number of galaxies is being carried out.  This 2D spectroscopy
provides the possibility to determine elemental abundances across
these objects and to construct abundance maps for the targeted
galaxies.  This permits one to investigate the abundance distribution
across the disc in detail. In particular, the  radial oxygen abundance
distributions in the discs of galaxies can be examined
\citep{Sanchez2014,Zinchenko2016}.  

A new calibration that provides estimates of the oxygen and nitrogen
abundances in star-forming regions with high precision over the whole
metallicity scale was recently suggested \citep{Pilyugin2016}.  The
mean difference between the calibration-based and $T_{e}$-based
abundances is around 0.05 dex both for oxygen and nitrogen.  The 2D
spectroscopic data mentioned above coupled with this new calibration
for nebular abundances can be used to explore the presence (or
absence) of breaks in the radial oxygen abundance distributions in 
galactic discs.

The goal of this investigation is to characterize the breaks in the
surface brightness and abundance distributions in a sample of
galaxies and to examine the relation between those breaks.

The paper is organized in the following way. The data are described in
Section 2.  The breaks of the radial abundance gradients are
determined in Section 3.  The breaks of the surface brightness
profiles and their relation to the breaks of the radial abundance
gradients are examined in Section 4. The discussion is given in
Section 5.  Section 6 contains a brief summary.

\section{Data}

\subsection{Spectral data}

We measured the different emission-line fluxes across galactic discs
using the 2D spectroscopy of galaxies carried out in the framework of
the CALIFA survey \citep{Sanchez2012, Husemann2013, GarciaBenito2015}.
We used publicly
available COMB data cubes from the CALIFA Data Release 3 (DR3), which combine spectra
obtained in high (V1200) and low (V500) resolution modes.  The
spectrum of each spaxel from the CALIFA datacubes was reduced in the
manner described in \citet{Zinchenko2016}. In brief, for each
spectrum, the fluxes of the 
[O\,{\sc ii}]$\lambda$3727+$\lambda$3729, 
H$\beta$,  
[O\,{\sc iii}]$\lambda$4959, 
[O\,{\sc iii}]$\lambda$5007,
[N\,{\sc ii}]$\lambda$6548,
H$\alpha$,  
[N\,{\sc ii}]$\lambda$6584, 
[S\,{\sc ii}]$\lambda$6717, and
[S\,{\sc ii}]$\lambda$6731 lines were measured.
The measured line fluxes were corrected for interstellar reddening
using the theoretical H$\alpha$ to H$\beta$ ratio  (i.e. the standard
value of H$\alpha$/H$\beta = 2.86$) and the analytical approximation
of the Whitford interstellar reddening law from \citet{Izotov1994}.
When the measured value of H$\alpha$/H$\beta$ was lower than 2.86 the
reddening was adopted to be zero. 

Since the [O\,{\sc iii}]$\lambda$5007 and $\lambda$4959 lines
originate from transitions from the same energy level, and their flux
ratio is constant, i.e. very close to 3 \citep{Storey2000}, and since the
stronger line, [O\,{\sc iii}]$\lambda$5007, is usually measured with
higher precision than the weaker line, [O\,{\sc iii}]$\lambda$4959, we
estimated the value of $R_3$ to be $R_3  = 1.33$~[O\,{\sc
iii}]$\lambda$5007/H$\beta$ but not as a sum of the line fluxes.  Similarly,
the [N\,{\sc ii}]$\lambda$6584 and $\lambda$6548 lines also originate
from transitions from the same energy level and the transition
probability ratio for those lines is again close to 3
\citep{Storey2000}. Therefore, we also estimated the value of $N_2$ to
be $N_2 = 1.33$~[N\,{\sc ii}]$\lambda$6584/H$\beta$. 

Thus, we used the lines 
H$\beta$,  
[O\,{\sc iii}]$\lambda$5007,
H$\alpha$,  
[N\,{\sc ii}]$\lambda$6584, 
[S\,{\sc ii}]$\lambda$6717, and
[S\,{\sc ii}]$\lambda$6731 
for the dereddening and abundance determinations. The
precision of the line flux is specified by the ratio of the flux to
the flux error $\epsilon$. We selected spectra for which the parameter
$\epsilon \geq 3$ for each of those lines. 
\citet{Belfiore2017} have also used the condition signal-to-noise ratio  
larger than three, S/N $>$ 3, for the
lines required to estimate abundances to select the spaxels for which
they derive abundances.

  We used a standard diagnostic diagram, [N\,{\sc
ii}]$\lambda$6584/H$\alpha$ versus\  [O\,{\sc
iii}]$\lambda$5007/H$\beta$ line ratios, suggested by
\citet{Baldwin1981}, which is known as the BPT classification diagram, to separate H\,{\sc ii} region-like objects and  AGN-like objects. We adopted the
demarcation line of \citet{Kauffmann2003} between H\,{\sc ii} regions
and AGNs. 

We obtained oxygen abundance maps for $\sim$250 spiral  
galaxies from CALIFA where the emission lines can be measured in a large 
number (typically more than a hundred) of spaxel spectra. 
Interacting galaxies are not taken in consideration. 
Our study is devoted to the examination of the presence of a break in the
radial abundance distribution in spiral galaxies.  We carry out a
visual inspection of the O/H -- $R_{g}$ diagram for each galaxy and
select a subsample of galaxies where the regions with measured oxygen
abundances are well distributed along the radius. Unfortunately, we
had to reject a number of galaxies whose abundance maps include a
large number of data points but where those points cover only a limited
fraction of galactocentric distance.  For example, the abundance maps
of the galaxies NGC~941, NGC~991, NGC~2602, and NGC~3395 contain more than
2000 data points each but those points are located within
galactocentric distances of less than $\sim 0.7 R_{25}$ and there are
no data points at distances larger than $\sim 0.7 R_{25}$.  The list
of galaxies selected for the current study includes 134 spiral
galaxies. 

The adopted and derived characteristics of our sample of galaxies 
are available in an online table.

\begin{figure*}
\resizebox{1.00\hsize}{!}{\includegraphics[angle=000]{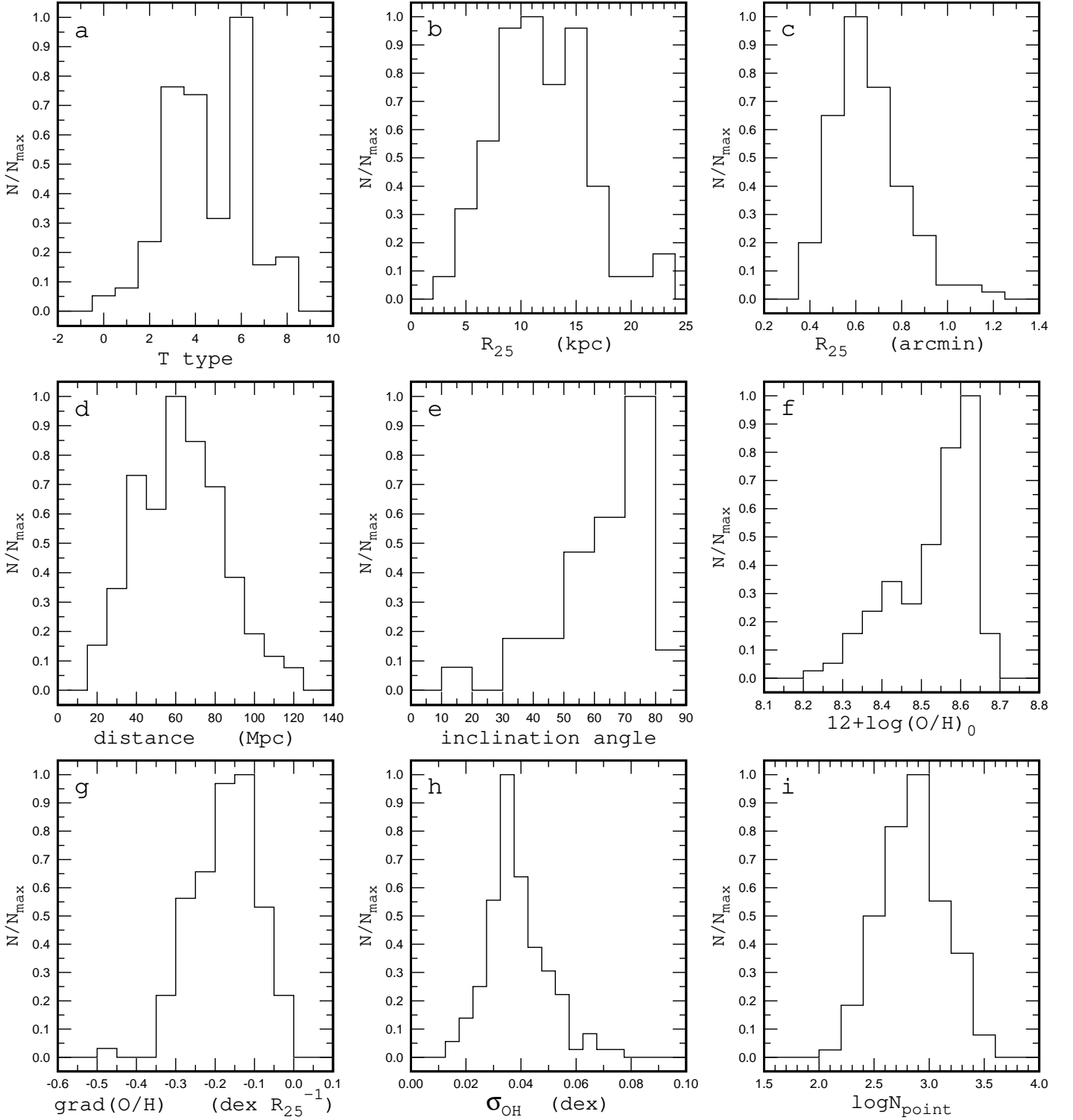}}
\caption{
Properties of our sample of galaxies. The panels show the 
normalized histograms of
morphological $T$ types ($a$),
optical radii $R_{25}$ in kpc ($b$),
optical radii $R_{25}$ in arcmin ($c$),
distances to the galaxies in Mpc ($d$),
galaxy inclination angles ($e$),
central (intersect) oxygen abundances 12+log(O/H)$_{0}$ ($f$),
radial oxygen abundance gradients in dex $R_{25}^{-1}$ ($g$),
scatter in oxygen abundances around the linear O/H -- $R_{g}$ 
relation ($h$), and
number of points in the abundance map of a given galaxy ($i$).
}
\label{figure:general}
\end{figure*}

Fig.~\ref{figure:general} shows the properties of our selected sample 
of galaxies, i.e. the normalized histograms of
morphological $T$ types (panel $a$),
optical radii $R_{25}$ in kpc (panel $b$),
optical radii $R_{25}$ in arcmin (panel $c$),
distances to our galaxies in Mpc (panel $d$),
galaxy inclination angles (panel $e$),
central (intersect) oxygen abundances 12+log(O/H)$_{0}$ (panel $f$),
radial abundance gradients in dex $R_{25}^{-1}$ (panel $g$),
scatter in oxygen abundance around the linear O/H -- $R_{g}$ 
relation (panel $h$), and
number of points in the abundance map (panel $i$).
The morphological classification and morphological $T$ types were
taken from the HyperLeda\footnote{http://leda.univ-lyon1.fr/} database
\citep{Paturel2003,Makarov2014}.  The distances were taken from the
NASA Extragalactic Database ({\sc ned})\footnote{The NASA/IPAC
Extragalactic Database ({\sc ned}) is operated by the Jet Propulsion
Laboratory, California Institute of Technology, under contract with
the National Aeronautics and Space Administration.  {\tt
http://ned.ipac.caltech.edu/} }.  The {\sc ned} distances use flow
corrections for Virgo, the Great Attractor, and Shapley Supercluster
infall.  Other parameters were derived in our current study (see
below).  Inspection of Fig.~\ref{figure:general} shows that the
selected galaxies are located at distances from $\sim 20$ to $\sim
120$ Mpc, belong to different morphological types, and show a large
variety of physical characteristics. The optical radii of the galaxies
cover the interval from around 4 kpc (UGC~5976, UGC~10803, UGC~12056)
to around 23 kpc (NGC~1324, NGC~6478, UGC~12810).  The abundance maps
typically contain hundreds and sometimes even a few thousand data
points.

\subsection{Photometry: Galaxy orientation parameters and isophotal radius}

To estimate the deprojected galactocentric distances (normalized to the
optical isophotal radius $R_{25}$) of the H\,{\sc ii} regions, we need
to know the galaxy inclinations, $i$, position angle of their major
axes, {\it PA}, and isophotal radius of our galaxies, $R_{25}$. 
Therefore we determined the values of $i$, {\it PA}, and
$R_{25}$ for our target galaxies by analysing publicly available
photometric images in the $g$ and $r$ bands obtained by the {\it Sloan
Digital Sky Survey} \citep[{\it SDSS}; data release 9
(DR9),][]{York2000,Ahn2012}.  We derived the surface brightness
profile and disc orientation parameters in these two photometric bands
for each galaxy.  

The determinations of the position angle and ellipticity were
performed for SDSS images in the $r$ band via GALFIT
\citep{Peng2010}. We fitted surface brightness proflies using a
two-component model. The bulge was fitted by a Sersic profile and the
disc by an exponential profile. In cases where the contribution of one
of the components to the total surface brightness profile was
significantly smaller the surface brightness profile of the dominant
component was refitted with only the major component profile.  The values
of the ellipticity and PA obtained for the exponential profile were
then adopted as galaxy ellipticity and PA. For galaxies without
exponential component the ellipticity and PA of the Sersic profile were
adopted instead as their ellipticity and PA.  The galaxies
NGC6090, NGC7549, UGC4425, and UGC10796, which show a peculiar surface
brightness distribution were considered as face-on galaxies.

The  {\it SDSS} photometry in different bands is sufficiently deep to
extend our surface brightness profiles well beyond the optical
isophotal radii $R_{25}$.  The value of the isophotal radius was
determined from the constructed surface brightness profiles in the $g$
and $r$ bands.  Our measurements were corrected for Galactic
foreground extinction using the recalibrated $A_V$ values of
\citep{Schlafly2011} taken from the {\sc
ned}.  Then the measurements in the {\it SDSS} filters $g$ and $r$
were converted to $B$-band magnitudes using the conversion relations
of \citet{Blanton2007},
\begin{equation} 
B_{AB} = g + 0.2354 + 0.3915\;[(g-r)-0.6102] ,
\label{equation:bgr} 
\end{equation} 
where the $B_{AB}$, $g$, and $r$ magnitudes in
Eq.~(\ref{equation:bgr}) are in the $AB$ photometric system.  The $AB$
magnitudes were reduced to the Vega photometric system 
\begin{equation} 
B_{Vega} = B_{AB} + 0.09  
\label{equation:VegaAB} 
\end{equation} 
using the relation of \citet{Blanton2007}.  The disc surface brightness 
in the $B$ band of the Vega photometric system reduced to the face-on position is  
used in the determination the optical isophotal radius $R_{25}$.

\section{Abundance gradients}

\subsection{Abundance determination}

It is believed that the classic $T_{e}$ method \citep[e.g.][]{Dinerstein1990}
provides the most reliable gas-phase abundance estimations in 
star-forming regions. However, abundance determinations through the
direct $T_{e}$ method require high-quality spectra in order
to measure temperature-sensitive auroral lines such as [O\,{\sc iii}]$\lambda$4363
or/and [N\,{\sc ii}]$\lambda$5755.  Unfortunately, these weak auroral lines
are not detected in the spaxel spectra of the CALIFA survey. 
But the abundances can be estimated through the strong
line methods first suggested by \citet{Pagel1979} and \citet{Alloin1979}.
We used the recent calibrations by \citet{Pilyugin2016},
which produce very reliable abundances as compared to other published calibrations.
Two variants ($R$ and $S$) of the calibration relations were suggested,
which provide abundances in close agreement with each other and with the $T_{e}$ method. The application of the $R$ calibration requires
the measurement of the oxygen emission lines [O\,{\sc ii}]$\lambda$3727+$\lambda$3729.
The sulphur emission lines [S\,{\sc ii}]$\lambda$6717+$\lambda$6731
are used instead of the oxygen lines [O\,{\sc ii}]$\lambda$3727+$\lambda$3729
in the $S$ calibration relations.  Since the sulphur emission
lines [S\,{\sc ii}]$\lambda$6717+$\lambda$6731 are more easily measured in
the spaxel spectra and with lower uncertainties than the oxygen emission lines  
[O\,{\sc ii}]$\lambda$3727+$\lambda$3729, we used the $S$ calibration
relations for the abundance determinations in the current study.

\begin{figure}
\resizebox{1.00\hsize}{!}{\includegraphics[angle=000]{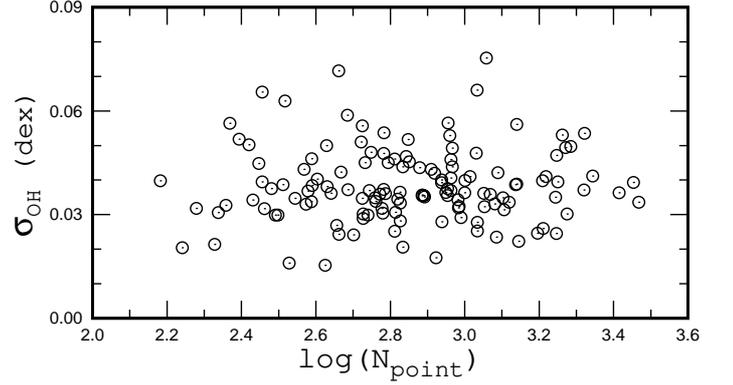}}
\caption{
Mean value of the residuals of the linear O/H -- $R_{g}$
relation as a function of the number of points in the abundance map of
our galaxies.
  }
\label{figure:npoint-sigma}
\end{figure}

\begin{figure*}
\resizebox{1.00\hsize}{!}{\includegraphics[angle=000]{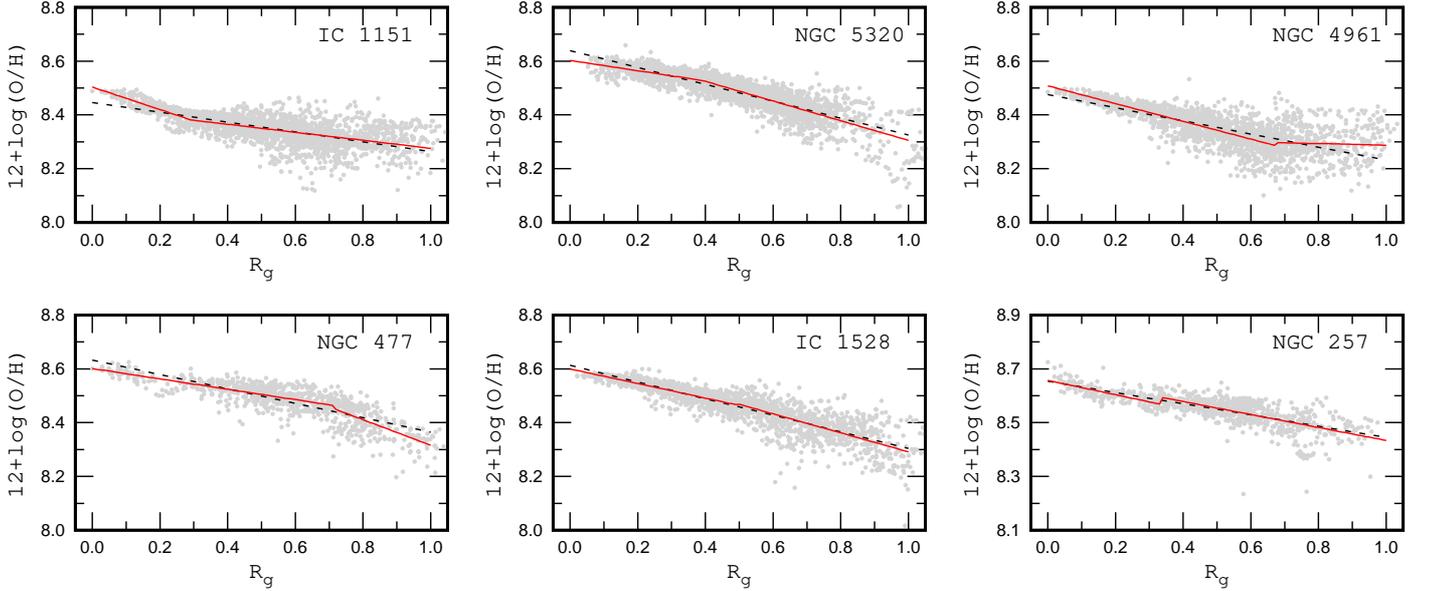}}
\caption{
Examples of the abundance distributions in the discs of spiral
galaxies.  The grey points depict the abundances in individual regions
(spaxels in the CALIFA survey).  The dashed (black) line shows the
pure linear fit to those data. The solid (red) line represents the
broken linear fit.  
}
\label{figure:galaxy-oh}
\end{figure*}

Simple calibration relations for abundances from a set of strong
emission lines were suggested  \citep{Pilyugin2016}.  Using
these relations, the oxygen abundances are determined using the
$N_2$ = 1.33[N\,{\sc ii}]$\lambda$6584/H$\beta$,
$R_3$ = 1.33[O\,{\sc iii}]$\lambda$5007/H$\beta$, and $S_2$
= ([S\,{\sc ii}]$\lambda$6717 +  [S\,{\sc ii}]$\lambda$6731)/H$\beta$ line
ratios  
\begin{eqnarray}
       \begin{array}{lll}
     {\rm (O/H)}^{*}  & = &   8.424 + 0.030 \, \log (R_{3}/S_{2}) + 0.751 \, \log N_{2}   \\  
                     & + &  (-0.349 + 0.182 \, \log (R_{3}/S_{2}) + 0.508 \log N_{2})   \\ 
                      & \times & \log S_{2}   \\ 
     \end{array}
\label{equation:ohsu}
\end{eqnarray}
if log$N_{2} > -0.6$, and  
\begin{eqnarray}
       \begin{array}{lll}
     {\rm (O/H)}^{*}  & = &   8.072 + 0.789 \, \log (R_{3}/S_{2}) + 0.726 \, \log N_{2}   \\  
                     & + &  (1.069 - 0.170 \, \log (R_{3}/S_{2}) + 0.022 \log N_{2})    \\ 
                     & \times & \log S_{2}   \\ 
     \end{array}
\label{equation:ohsl}
\end{eqnarray}
if log$N_{2} \le  -0.6$.  The notation (O/H)$^{*}$ = 12 +log(O/H)
is used for sake of brevity.

This calibration provides estimates of the oxygen abundances with high
precision over the whole metallicity scale. The mean difference
between the calibration-based and $T_{e}$-based oxygen abundances is
around 0.05 dex \citep{Pilyugin2016}. Thus our metallicity scale is
well compatible with the metallicity scale defined by the $T_{e}$-based
oxygen abundances.

\subsection{Fits to radial abundance distribution}

Two different fits to the radial oxygen abundance distribution are
considered. First, we fit the radial oxygen abundance distribution in
each galaxy by the traditional, purely linear relation, 
\begin{equation}
12 +  \log {\rm (O/H)}_{PLR} = 12 +  \log {\rm (O/H)}_{0} + grad \, \times \, R_{g}
\label{equation:1l}
,\end{equation}
where 12 + log(O/H)$_{0}$ is the extrapolated central oxygen
abundance, $grad$ is the slope of the oxygen abundance gradient
expressed in terms of dex/$R_{\rm 25}$, and $R_{g}$ is the fractional
radius (the galactocentric distance normalized to the disc isophotal
radius).

If there are points that show large deviations from the O/H --
$R_{g}$ relation, $d_{OH} > 0.3$ dex, then those points are not
used in deriving the final relations and are excluded from further
analysis.  It should be noted that points with such large deviations
exist only in some galaxies and those points are both few in
number and their deviations exceed 5 -- 10$\sigma_{OH}$ (see
below).  The mean deviation from the final relations (the mean value
of the residuals of the relations) is given by the expression  
\begin{equation}
\sigma_{OH} = \left(\frac{1}{n}\sum_{j=1}^{n} (\log ({\rm O/H})^{OBS}_{j} - \log ({\rm O/H})^{CAL}_{j})^{2} \right)^{1/2}  
\label{equation:delta}
,\end{equation}
where (O/H)$_{j}^{CAL}$ is the oxygen abundance computed through Eq.~(\ref{equation:1l}) 
for the galactocentric distance of the $j$-th spaxel and (O/H)$_{j}^{OBS}$ is the measured oxygen abundance
that is obtained through the calibration from the set of emission lines 
in the spectrum of the $j$-th spaxel.
The value of $\sigma_{OH}$ 
is usually low, from $\sigma_{OH} \sim0.03$ to $\sigma_{OH} \sim 0.06$
dex (see Fig.~\ref{figure:npoint-sigma}), and does not depend on the
number of data points in the abundance map of a galaxy.  This suggests
that the typical amplitude of the variation in the oxygen abundance at
a given galactocentric distance is rather small. 

We also fit the radial oxygen abundance distribution in each galaxy 
by a broken linear relation,
\begin{eqnarray}
       \begin{array}{lll}
12 + \log {\rm (O/H)}_{BLR}  & = &   a_{1} \, \times \, R_{g}  + b_{1}, \; \; \; \; \mbox{  if} \; \; R_{g} < R_{b}   \\  
                            & = &   a_{2} \, \times \, R_{g}  + b_{2}, \; \; \; \; \mbox{  if} \; \; R_{g} \ge R_{b}   \\  
     \end{array}
\label{equation:2l}
,\end{eqnarray}
where $R_{b}$ is the break radius, $b_{1}$ is the extrapolated central
oxygen abundance, $a_{1}$ is the slope of the oxygen abundance
gradient expressed in terms of dex/$R_{\rm 25}$ for the points in the
inner part of a galaxy, $b_{2}$ is the extrapolated central oxygen
abundance, and $a_{2}$  is the slope of the oxygen abundance gradient
for the point in the outer part of a galaxy.  The values of $R_{b}$,
$b_{1}$, $a_{1}$, $b_{2}$, and $a_{2}$ are derived using the requirement
that the mean deviations (Eq.~\ref{equation:delta}) from the relation
given by Eq.~\ref{equation:2l} are minimized. 

When determining the break in the radial abundance gradient at very
small galactocentric distances (near the centre of a galaxy) the
following problem may occur. The gradient based on a small range of
galactocentric distances or/and on a small number of points can be
unreliable. Even a few points with reduced (enhanced) abundances near
the centre can result in a (false) break if the (false) break radius
$R_{b}$ is small and the number of points within  $R_{b}$ is small.
The determination of the break in the radial abundance gradient at
very large galactocentric distances (near the optical radius $R_{25}$)
can also encounter the same problem. To avoid this problem we used the
additional requirements that both the inner and outer gradients
must be based on at least 50 data points that cover a range of
galactocentric distances larger than $0.2 R_{25}$. Those additional 
conditions may result in an underestimation or overestimation of the
slope of the inner (outer) gradients in some galaxies as well as shift
the position of the small break radius ($R_{b} < 0.2R_{25}$) to the
$R_{b} = 0.2R_{25}$ and the position of the large break radius ($R_{b}
> 0.8R_{25}$) to $R_{b} = 0.8R_{25}$. 

Fig.~\ref{figure:galaxy-oh} shows prototypical examples of abundance
distributions in galactic discs.  The grey points in each panel indicate
the abundances in individual regions (spaxels) estimated from the
CALIFA survey spectra.  The dashed (black) line depicts a purely
linear fit to those data while the solid (red) line shows a broken
linear fit to the data.

\subsection{On the validity of radial abundance distributions}

\begin{figure*}
\resizebox{0.95\hsize}{!}{\includegraphics[angle=000]{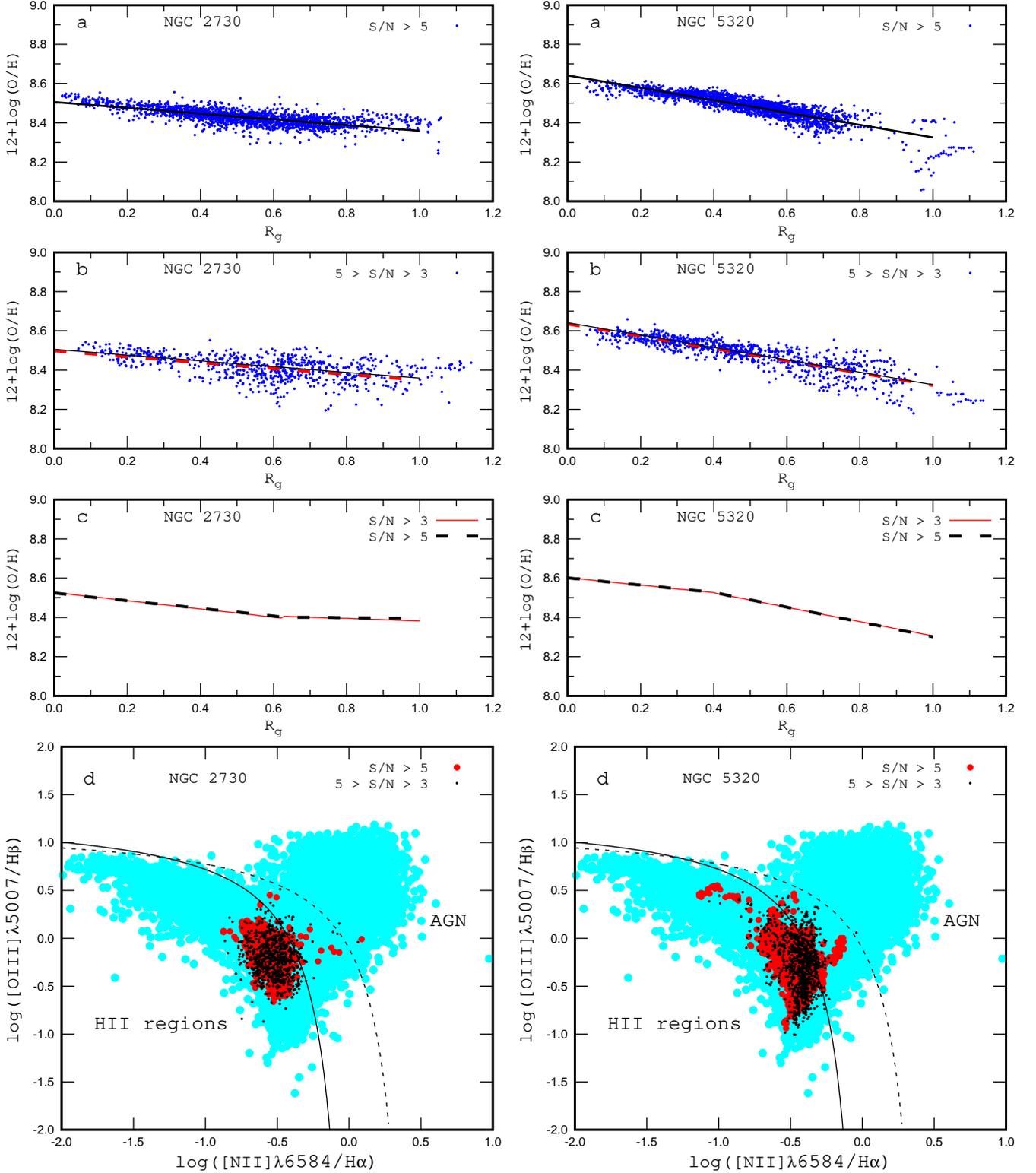}}
\caption{
Left column panel shows the properties of the galaxy NGC~2730.
Panel $a$ shows the radial abundance distributions traced by spaxels 
with high-quality spectra, i.e. with a S/N $>$ 5 for each of the measured emission lines.
The points depict individual spaxels. The dark (black) solid line is the linear best fit 
to those data.
Panel $b$ shows the radial abundance distributions traced by the spaxels 
with moderate-quality spectra, i.e. with a 5 $>$ S/N $>$ 3 for each of the measured emission lines.
The points denote individual spaxels, the dark grey (red) dashed line is 
the pure linear best fit to those data, and the dark (black) solid line 
is the same as in panel $a$. Panel $c$ shows the comparison between the 
broken linear fits to the data for the specta with S/N $>$ 3 (dark grey 
(red) solid line) and for the spectra with S/N $>$ 5 (dark (black) dashed line).  
Panel $d$ shows the classification [N\,{\sc ii}]$\lambda$6584/H$\alpha$ 
vs. [O\,{\sc iii}]$\lambda$5007/H$\beta$ diagram. 
The dark grey (red) points indicate the spaxels with high-quality spectra.
The dark (black) points stand for the spaxels with moderate-quality spectra. 
The solid line separates objects with H\,{\sc ii} spectra from those 
containing an AGN according to \citet{Kauffmann2003}, while the dashed line is the same boundary 
according to \citet{Kewley2001}. 
The grey (light blue) filled circles show a large sample of emission-line SDSS 
galaxies \citep{Thuan2010}. 
The right column panels show the same but for the galaxy NGC~5320.
}
\label{figure:sn5}
\end{figure*}

\begin{figure}
\resizebox{1.00\hsize}{!}{\includegraphics[angle=000]{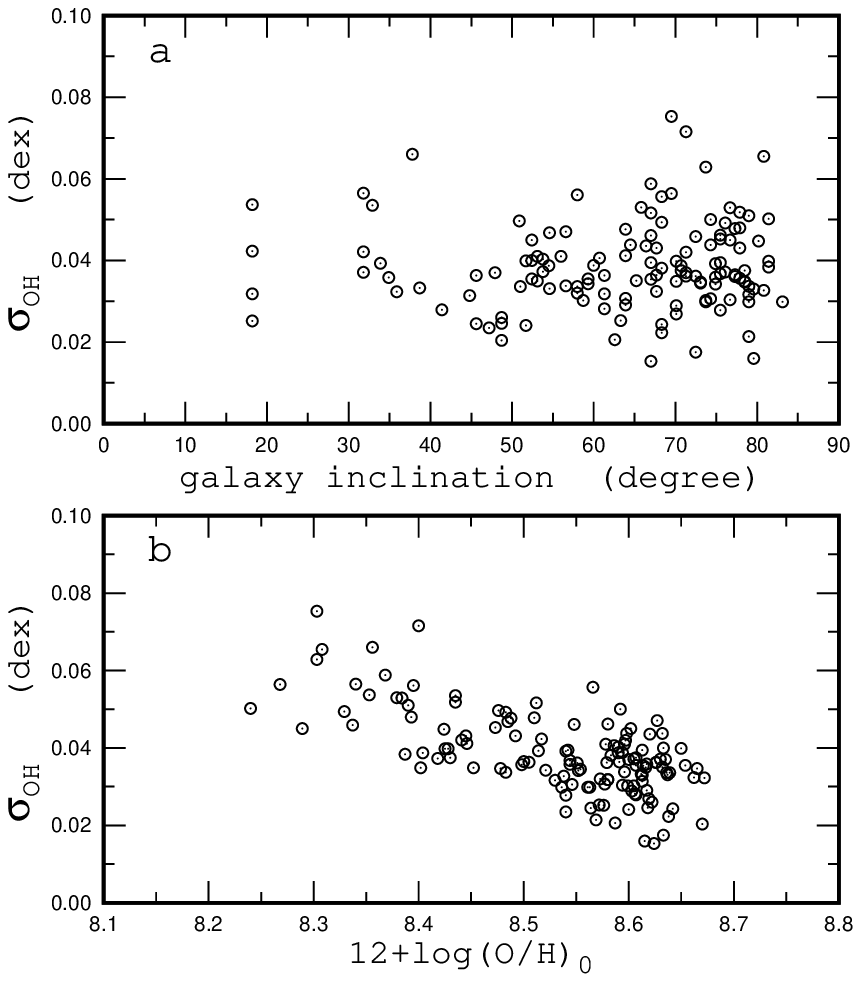}}
\caption{
Scatter of the oxygen abundances $\sigma_{OH,PLR}$ around the 
purely linear relation O/H -- $R_{g}$ as a function of galaxy 
inclination (upper panel) and central oxygen abundance (lower panel).
}
\label{figure:i-sigma}
\end{figure}

As noted above, we included all spaxel spectra in the analysis where the ratio of the flux to
the flux error, S/N $>$ 3, for each of the lines used in the abundance determinations.
If the number of the spaxel spectra with measured
emission lines in a given galaxy is large enough then the influence of the adopted level
of the S/N of the  line measurement on the parameters of the abundance distribution
can be investigated.

The left column panels of Fig.~\ref{figure:sn5} show the properties of the galaxy NGC~2730,
where the gradient in the outer part of the disc is flatter than that in the inner part. 
Panel $a$ shows the radial abundance distributions traced by spaxels 
with high-quality spectra, i.e. a S/N $>$ 5 for each of the measured emission lines.
The points stand for individual spaxels. The solid (black) line is the purely linear best fit 
to those data.
Panel $b$ shows the radial abundance distributions traced by spaxels 
with moderate-quality spectra, i.e. with a 5 $>$ S/N $>$ 3 for each of the measured emission lines. 
The points denote individual spaxels, the dashed (red) line is 
the purely linear best fit to those data, and the solid (black) line 
is the same as in the panel $a$. Panel $c$ shows the comparison between the 
broken linear fits to the data for specta with S/N $>$ 3 (solid  
(red) line) and for spectra with S/N $>$ 5 (dashed (black) line).
There is a remarkable agreement between those fits.
Panel $d$ shows the [N\,{\sc ii}]$\lambda$6584/H$\alpha$ 
versus [O\,{\sc iii}]$\lambda$5007/H$\beta$ classification diagram. 
The red points indicate spaxels with high-quality spectra.
The black points indicate spaxels with moderate-quality spectra. 
The solid line separates objects with H\,{\sc ii} region spectra from those 
containing an AGN according to \citet{Kauffmann2003}, while the dashed line is the boundary line 
according to \citet{Kewley2001}. 
The light blue points show a large sample of emission-line SDSS 
galaxies \citep{Thuan2010}. 
The right column panels of Fig.~\ref{figure:sn5} show the same but for the galaxy NGC~5320,
where the gradient in the inner part of the disc is flatter than that in the outer part.

Inspection of panels $a$ and $b$ of Fig.~\ref{figure:sn5} shows
that both the spaxels with high-quality spectra as well as the spaxels with moderate-quality spectra are distributed across the whole extent of the
discs of both galaxies.  Panels $d$ in Fig.~\ref{figure:sn5} show 
that the spaxels with moderate-quality spectra occupy the same area in the
diagnostic [N\,{\sc ii}]$\lambda$6584/H$\alpha$ versus [O\,{\sc iii}]$\lambda$5007/H$\beta$
diagram as the spaxels with high-quality spectra.

A comparison between the left column panels $a$ and $b$ of Fig.~\ref{figure:sn5}
shows that the purely linear gradient in the disc of the galaxy NGC~2730 traced
by spaxels with high-quality spectra (1883 data points)
is in agreement with the purely linear gradient traced by the spaxels with
moderate-quality spectra (718 data points). However, the scatter of the abundances
in spaxels with high-quality spectra around the O/H -- $R_{g}$ relation,
$\sigma_{\rm OH,PLR} = 0.0283$, is lower than that for the spaxels with moderate-quality spectra, $\sigma_{\rm OH,PLR} = 0.0510$.
A similar behaviour is seen in the
galaxy NGC~5320 (right column panels $a$ and $b$ of Fig.~\ref{figure:sn5}).
While the purely linear gradients traced by the spaxels with high-quality spectra (2045 data points)
and by the spaxels with moderate-quality spectra (890 data points) are
in agreement, the scatter of the abundances
in spaxels with high-quality spectra around the O/H -- $R_{g}$ relation,
$\sigma_{\rm OH,PLR}$ = 0.0297, is lower than that for the spaxels with moderate-quality spectra, $\sigma_{\rm OH,PLR}$ = 0.0410.
The left and right column panels $c$ show that the broken linear O/H -- $R_{g}$ relation
traced by the spaxels with high-quality spectra agrees with relation 
traced by all the spaxels (i.e. the spaxels with high-quality spectra and spaxels with
moderate-quality spectra taken together).

Thus, the use of spaxel spectra with a 
S/N $>$ 3 for each of the lines used in the abundance determinations
is justified and does not influence the derived characteristics of the abundance
distribution, i.e. the slope of the gradient and the break in the abundance distribution,
although it increases the scatter around the  O/H -- $R_{g}$ relation.
This result is not surprising since the condition S/N $>$ 3 is the
standard criterion and has been widely used for a long time.
 
Our sample involves a number of galaxies with large inclinations, 
$i > 70{\degr}$ (see panel $e$ in Fig.~\ref{figure:general}). 
One can expect that the deprojected galactocentric distances of the 
spaxels in galaxies with large inclinations 
may suffer from large uncertainties that can result in an increase of 
the scatter around the O/H -- $R_{g}$ relation. 
Fig.~\ref{figure:i-sigma} shows the scatter of the oxygen abundances $\sigma_{OH,PLR}$ 
around the purely linear relation O/H -- $R_{g}$ as a function of galaxy 
inclination (upper panel) and central oxygen abundance (lower panel).
The upper panel of Fig.~\ref{figure:i-sigma} shows that the value of 
the scatter $\sigma_{OH,PLR}$ is not correlated with 
the inclination. This suggests that a reliable  O/H -- $R_{g}$ 
diagram can be obtained even for galaxies with a large inclination, 
up to $i \sim 80{\degr}$.

Inspection of the lower panel of Fig.~\ref{figure:i-sigma} shows that 
there is a correlation between the value of the oxygen abundance scatter 
$\sigma_{OH,PLR}$ and the (central) oxygen abundance in the sense 
that the scatter increases with decreasing oxygen abundance. 
This could have the following cause. The flux in the 
nitrogen emission line [N\,{\sc ii}]$\lambda$6584, which plays a significant role 
in the abundance determinations decreases with decreasing galaxy 
metallicity. This can result in an increase of the uncertainties in the 
measurements of the [N\,{\sc ii}]$\lambda$6584 emission-line fluxes 
with decreasing galaxy metallicity and, as a consequence, 
may lead to an increase of the oxygen abundance scatter with 
decreasing abundance. 
It also cannot be excluded that the real scatter in the abundances is large 
in the less evolved (lower abundance and astration level) galaxies than in 
the more evolved (higher abundance and astration level) galaxies. 
If this is the case then this effect may contribute to the 
observed trend $\sigma_{OH,PLR}$ -- O/H.

\subsection{Properties of radial abundance distributions}

\begin{figure}
\resizebox{1.00\hsize}{!}{\includegraphics[angle=000]{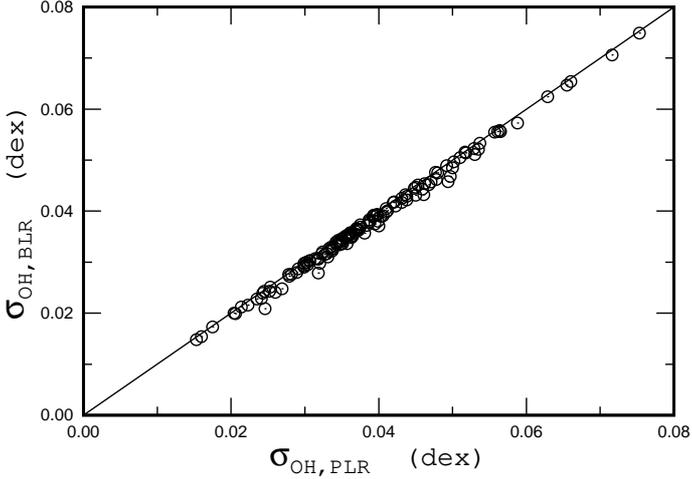}}
\caption{Scatter of oxygen abundances $\sigma_{OH,BLR}$
around broken linear relation O/H -- $R_{g}$ vs. scatter
$\sigma_{OH,PLR}$ around the purely linear relation. The dashed line 
indicates unity. 
}
\label{figure:sohp-sohb}
\end{figure}

\begin{figure}
\resizebox{1.00\hsize}{!}{\includegraphics[angle=000]{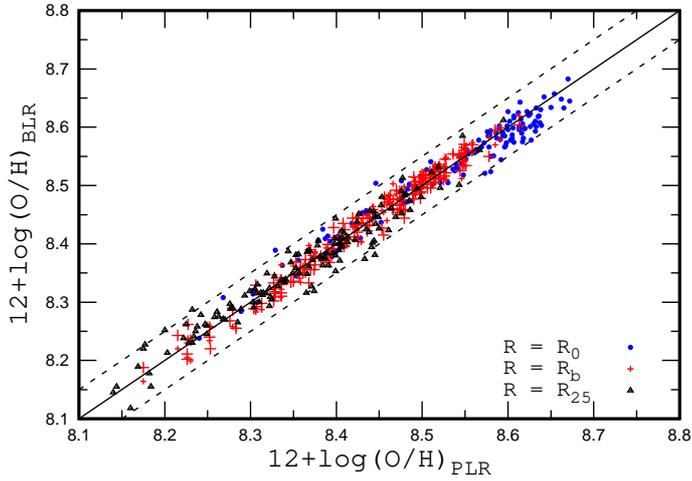}}
\caption{
Comparison between oxygen abundances (O/H)$_{PLR}$  given by the
purely linear relation (O/H) -- $R_{g}$ and abundances
(O/H)$_{BLR}$ given by the broken linear relation at the galaxy centres
(circles), at the break radii (plus signs), and at the optical radii
$R_{25}$ (triangles) of the galaxies.  The solid line indicates equal
values. The two dashed lines show the $\pm 0.05$ dex deviations from
unity.
}
\label{figure:ohp-ohb}
\end{figure}

\begin{figure}
\resizebox{1.00\hsize}{!}{\includegraphics[angle=000]{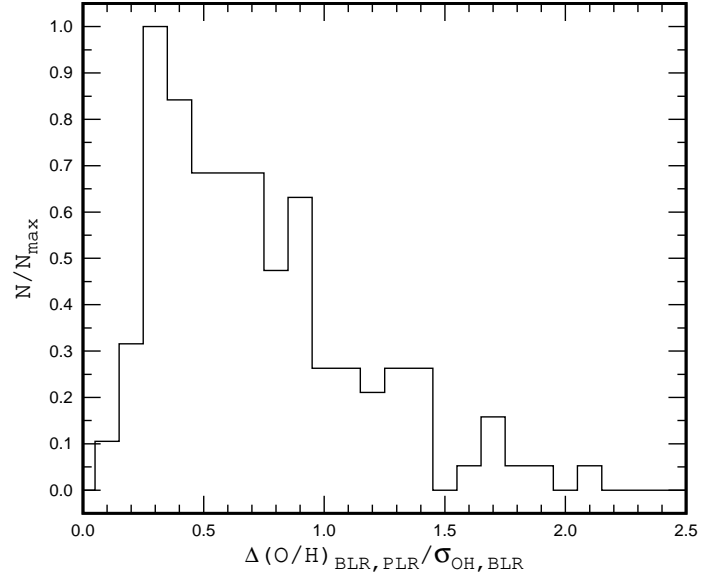}}
\caption{ 
Normalized histogram of the ratios of the difference between broken
and linear relation  O/H -- $R_{g}$  to the abundance scatter around the broken
relation $\Delta$(O/H)$_{BLR,PLR}$/$\sigma_{OH,BLR}$ for our sample of galaxies. 
}
\label{figure:gist-doh}
\end{figure}

\begin{table}
\caption{Break radii in  $R_{g}$ -- (O/H)$_{X}$ relations for oxygen abundances 
(O/H)$_{X}$ determined through different calibrations: the (O/H)$_{\rm PG16}$ abundances were determined 
using the $S$ calibration of \citet{Pilyugin2016}, whereas the (O/H)$_{\rm M13,O3N2}$ and (O/H)$_{\rm M13,N2}$
abundances were obtained through the O3N2 and N2 calibrations of \citet{Marino2013}, respectively. 
The break radii are fractional radii, i.e. they are normalized to the optical radii. 
}
\label{rbrb}
\begin{tabular}{lccc} 
\hline    
                    & \multicolumn{3}{c}{break radius in $R_{g}$ -- (O/H)$_{X}$ relation} \\
\cline{2-4}                
 Galaxy             & (O/H)$_{\rm PG16}$ & (O/H)$_{\rm M13,O3N2}$ &  (O/H)$_{\rm M13,N2}$   \\   \hline
NGC~0477            &   0.71           &   0.35             &     0.69             \\
NGC~0768            &   0.55           &   0.55             &     0.39             \\
NGC~3811            &   0.46           &   0.47             &     0.34             \\
NGC~4644            &   0.80           &   0.48             &     0.80             \\
NGC~4961            &   0.67           &   0.57             &     0.61             \\
NGC~5320            &   0.40           &   0.41             &     0.50             \\
NGC~5630            &   0.53           &   0.53             &     0.54             \\
NGC~5633            &   0.47           &   0.27             &     0.55             \\
NGC~5980            &   0.77           &   0.49             &     0.80             \\
NGC~6063            &   0.75           &   0.40             &     0.77             \\
NGC~6186            &   0.40           &   0.58             &     0.37             \\
NGC~6478            &   0.76           &   0.47             &     0.70             \\
NGC~7364            &   0.42           &   0.20             &     0.24             \\
NGC~7489            &   0.33           &   0.31             &     0.28             \\
NGC~7631            &   0.67           &   0.55             &     0.45             \\
IC~0480             &   0.76           &   0.32             &     0.76             \\
IC~1151             &   0.76           &   0.28             &     0.79             \\
IC~2098             &   0.78           &   0.50             &     0.50             \\
UGC~00005           &   0.44           &   0.29             &     0.64             \\            
UGC~00841           &   0.34           &   0.32             &     0.60             \\            
UGC~01938           &   0.42           &   0.48             &     0.42             \\            
UGC~02319           &   0.79           &   0.80             &     0.79             \\            
UGC~04029           &   0.73           &   0.60             &     0.79             \\            
UGC~04132           &   0.79           &   0.38             &     0.53             \\            
UGC~09665           &   0.74           &   0.70             &     0.80             \\            
UGC~12857           &   0.80           &   0.80             &     0.80             \\            
\hline
\end{tabular}
\end{table}

It is evident that any radial abundance distribution is fitted better
by the broken linear relation than by the purely linear relation. The
general difference between pure and broken linear relations can be
specified by the difference of the scatter of the abundances
around those relations and by the maximum difference between the abundances given by those
relations.

Fig.~\ref{figure:sohp-sohb} shows the scatter of the oxygen abundances
$\sigma_{OH,BLR}$ around the broken linear relation O/H -- $R_{g}$
versus the scatter $\sigma_{OH,PLR}$ around the purely linear
relation.  The circles show individual galaxies. The dashed line is
that of equal values.  Fig.~\ref{figure:sohp-sohb} illustrates
that the scatter of the oxygen abundances $\sigma_{OH,BLR}$ is very
close to the scatter $\sigma_{OH,PLR}$. The difference between those
values is negligibly small in the majority of our galaxies.

The difference between the broken and purely linear (O/H) -- $R_{g}$
 relations can be specified in the following way. 
The difference between the abundance (O/H)$_{BLR}$
corresponding to the broken (O/H) -- $R_{g}$ relation and the
abundance (O/H)$_{PLR}$ corresponding to the purely linear (O/H) --
$R_{g}$ relation is defined as $\Delta$(O/H)$_{BLR,PLR}$($R_{g}$) = log(O/H)$_{BLR}$($R_{g}$) -
log(O/H)$_{PLR}$($R_{g}$) and changes in radial direction. The maximum difference
occurs either at the centre of a galaxy, break radius $R_{b}$, or optical isophotal radius $R_{25}$.

Fig.~\ref{figure:ohp-ohb} shows the oxygen abundance (O/H)$_{BLR}$ as
a function of the abundance (O/H)$_{PLR}$ at the centres (circles), break radii (plus signs), and optical radii (triangles) of
our galaxies. Since there may be a jump in oxygen abundance at the
break radius we are presenting two values of the (O/H)$_{BLR}$ at the
break radius estimated from the gradients of the inner and outer parts
of the disc.  The solid line is that of equal values; the dashed lines
show the $\pm$0.05 dex deviations from unity.  Inspection of
Fig.~\ref{figure:ohp-ohb} shows that the maximum difference between
the abundances (O/H)$_{BLR}$ and (O/H)$_{PLR}$ is within 0.05 dex for
the majority of the galaxies of our sample. 
 
We specify the difference between the broken and purely linear (O/H) -- $R_{g}$
relations by the maximum absolute value of the difference. This value is referred 
to as $\Delta$(O/H)$_{BLR,PLR}$ below.  
Fig.~\ref{figure:gist-doh} shows the normalized histogram of the ratios of the
difference between the broken and linear the relation O/H -- $R_{g}$ to the abundance
scatter around the broken relation $\Delta$(O/H)$_{BLR,PLR}$/$\sigma_{OH,BLR}$
for our galaxy sample. 
Inspection of Fig.~\ref{figure:gist-doh} shows that 
the difference between the broken and linear relations exceeds 
the abundance scatter around the broken relation for 26 galaxies out of 134 objects.

\subsection{Breaks in the radial distributions of abundances obtained through various calibrations}

\begin{figure*}
\resizebox{1.00\hsize}{!}{\includegraphics[angle=000]{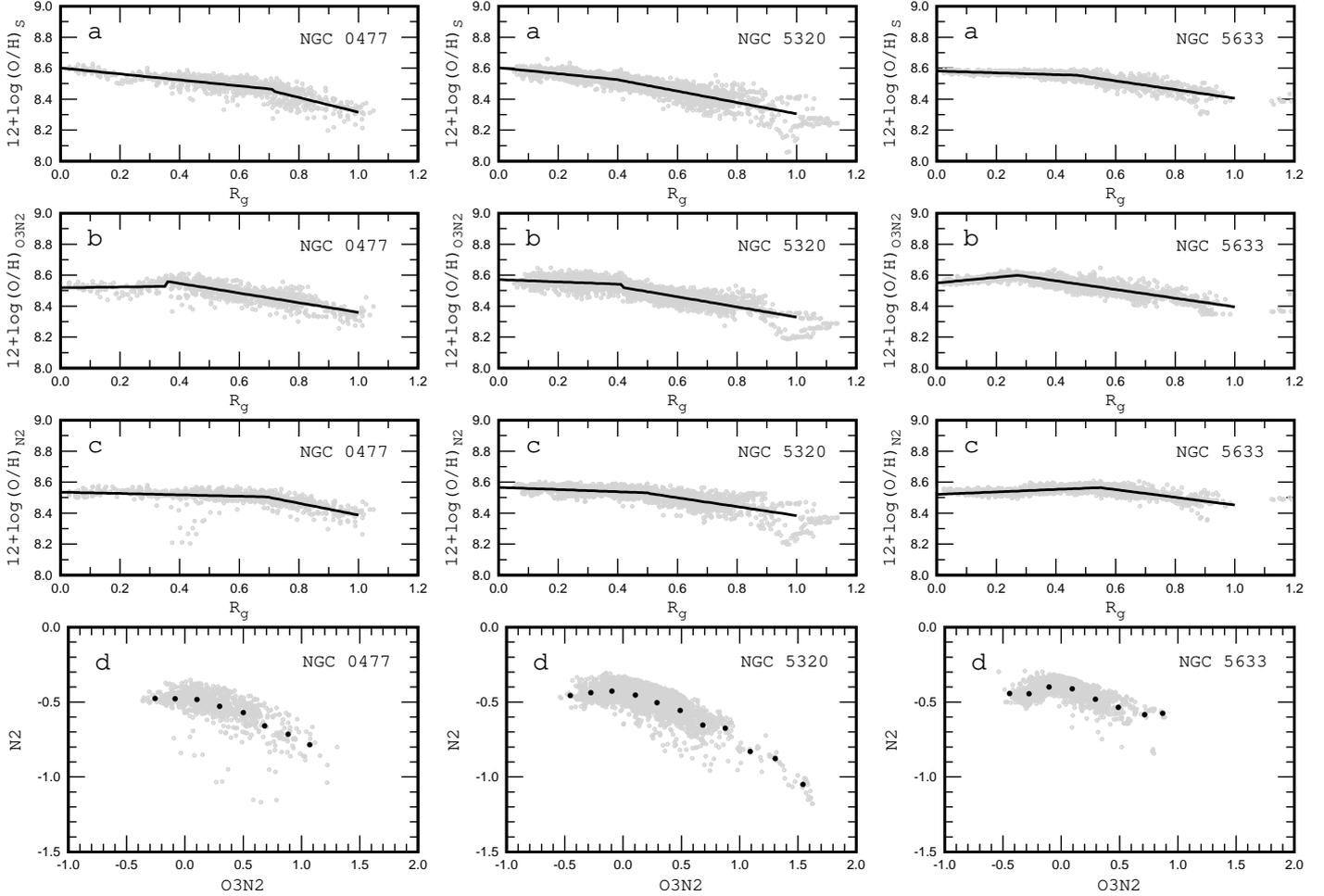}}
\caption{
Breaks in the radial distributions of oxygen abundances in discs of three galaxies
determined via three different calibrations are depicted. 
The left column panels show the galaxy NGC~0477, where the break radii in the (O/H)$_{\rm PG16}$
and the (O/H)$_{\rm M13,N2}$ distributions are close to each other and differ from the break radius obtained from 
the (O/H)$_{\rm M13,O3N2}$ distribution. The grey points in panels $a$, $b$, and $c$ are the data for
individual spaxels. The line is the broken linear fit O/H -- $R_{g}$ to those data points.
Panel $d$ shows the N2 abundance index as a function of the O3N2 index for individual spaxels
(grey points) and mean values for bin sizes of 0.2 in O3N2 (dark points).
The middle column panels show the galaxy NGC~5320, where the break radii in the (O/H)$_{\rm PG16}$
and the (O/H)$_{\rm M13,O3N2}$ distributions are close to each other and differ from that in  
the (O/H)$_{\rm M13,N2}$ distribution.
The right column panels show the galaxy NGC~5633, where the break radii for the three
abundance determination methods are different.
}
\label{figure:example-rpg-rma}
\end{figure*}

The 1D O3N2 and N2 calibrations suggested by \citet{Pettini2004} are widely used
for abundance estimations. We aim to find out whether the radial distributions of abundances obtained with
those calibrations show breaks in the slopes of their gradients.
We found above that the radial distributions of oxygen abundances estimated through
the calibration of \citet{Pilyugin2016} show a prominent break in the abundance gradients
in the discs of 26 galaxies. 
The oxygen abundances in those galaxies were determined 
using the recent version of the Pettini \& Pagel calibrations suggested by \citet{Marino2013}.
The (O/H)$_{\rm M13,O3N2}$ and (O/H)$_{\rm M13,N2}$ abundances were determined using the calibration relations 
\begin{equation} 
12+\log({\rm O/H})_{\rm M13,O3N2} = 8.533 - 0.214 \times {\rm O3N2}
\label{equation:OHo3n2} 
,\end{equation} 
where
\begin{equation} 
{\rm O3N2}  = \log\left(\frac{[{\rm O~III}]\lambda5007}{{\rm H}\beta}/\frac{[{\rm N~II}]\lambda6584}{{\rm H}\alpha}\right)
\label{equation:o3n2} 
\end{equation} 
and
\begin{equation} 
12+\log({\rm O/H})_{\rm M13,N2} = 8.743 + 0.462 \times {\rm N2}
\label{equation:OHn2} 
,\end{equation} 
where
\begin{equation} 
{\rm N2}  = \log\left(\frac{[{\rm N~II}]\lambda6584}{{\rm H}\alpha}\right)
\label{equation:n2} 
.\end{equation} 
The obtained radial abundance distribution in each galaxy is fitted by a broken linear relation.
The break radii in the O/H -- $R_{g}$ relations for oxygen abundances 
determined through the three different calibrations  are reported in Table~\ref{rbrb}; the abundances determined
through the $S$ calibration of \citet{Pilyugin2016} are referred 
to as (O/H)$_{\rm PG16}$ abundances in this section.

Table~\ref{rbrb} shows that the break radii in the
(O/H)$_{\rm PG16}$ -- $R_{g}$ and (O/H)$_{\rm M13,O3N2}$ -- $R_{g}$ distributions
agree within 0.1$R_{25}$ for 11 galaxies, the break radii in the
(O/H)$_{\rm PG16}$ -- $R_{g}$ and (O/H)$_{\rm M13,N2}$ -- $R_{g}$ distributions
agree for 18 galaxies, and the break radii 
in (O/H)$_{\rm M13,O3N2}$ -- $R_{g}$ and (O/H)$_{\rm M13,N2}$ -- $R_{g}$ distributions
agree for 10 galaxies. 
Panels $a$, $b$, and $c$ in Fig.~\ref{figure:example-rpg-rma} show the breaks in the radial
distributions of oxygen abundances in discs of three galaxies determined through
three different calibrations. The left column panels show the galaxy NGC~0477 where
the break radii in the (O/H)$_{\rm PG16}$ and (O/H)$_{\rm M13,N2}$ distributions are
close to each other, but differ from that in the (O/H)$_{\rm M13,O3N2}$ distribution. 
The middle column panels show the galaxy NGC~5320 where the break radii in the (O/H)$_{\rm PG16}$
and (O/H)$_{\rm M13,O3N2}$ distributions are close to each other, but differ from that in  
the (O/H)$_{\rm M13,N2}$ distribution.  The right column panels show the galaxy NGC~5633 where
the break radii in the three distributions are different.

Panels $d$ in Fig.~\ref{figure:example-rpg-rma} show the N2 abundance index as a
function of the O3N2 index for individual spaxels (grey points) and mean values
with a bin size of 0.2 in O3N2 (dark points) for the three galaxies. 
The N2 -- O3N2 diagrams for the individual galaxies
(panels $d$ in Fig.~\ref{figure:example-rpg-rma}) suggest that the 
disagreement between the break radii in the O/H -- $R_{g}$ relations for oxygen abundances originates in the use of 
the three different calibrations. Indeed, the relation between O3N2 and the
N2 abundance indicators is not strictly linear. Instead there is an appreciable deviation
from the linear relation. Hence, the 1D linear expressions O/H = $f$(O3N2)
and  O/H = $f$(N2) cannot both be correct simultaneously and cannot provide consistent 
(correct) values of the oxygen abundances over whole interval of the abundance index O3N2
(or the abundance index N2). This may result in a false apparent break
in the distribution of abundances produced by one (or both) of the 1D calibrations
of the Pettini \& Pagel type and, consequently, may lead to the disagreement in the 
break radii in the distributions of the abundances produced by the O3N2 and N2 calibrations.

\begin{figure}
\begin{center}
\resizebox{0.98\hsize}{!}{\includegraphics[angle=000]{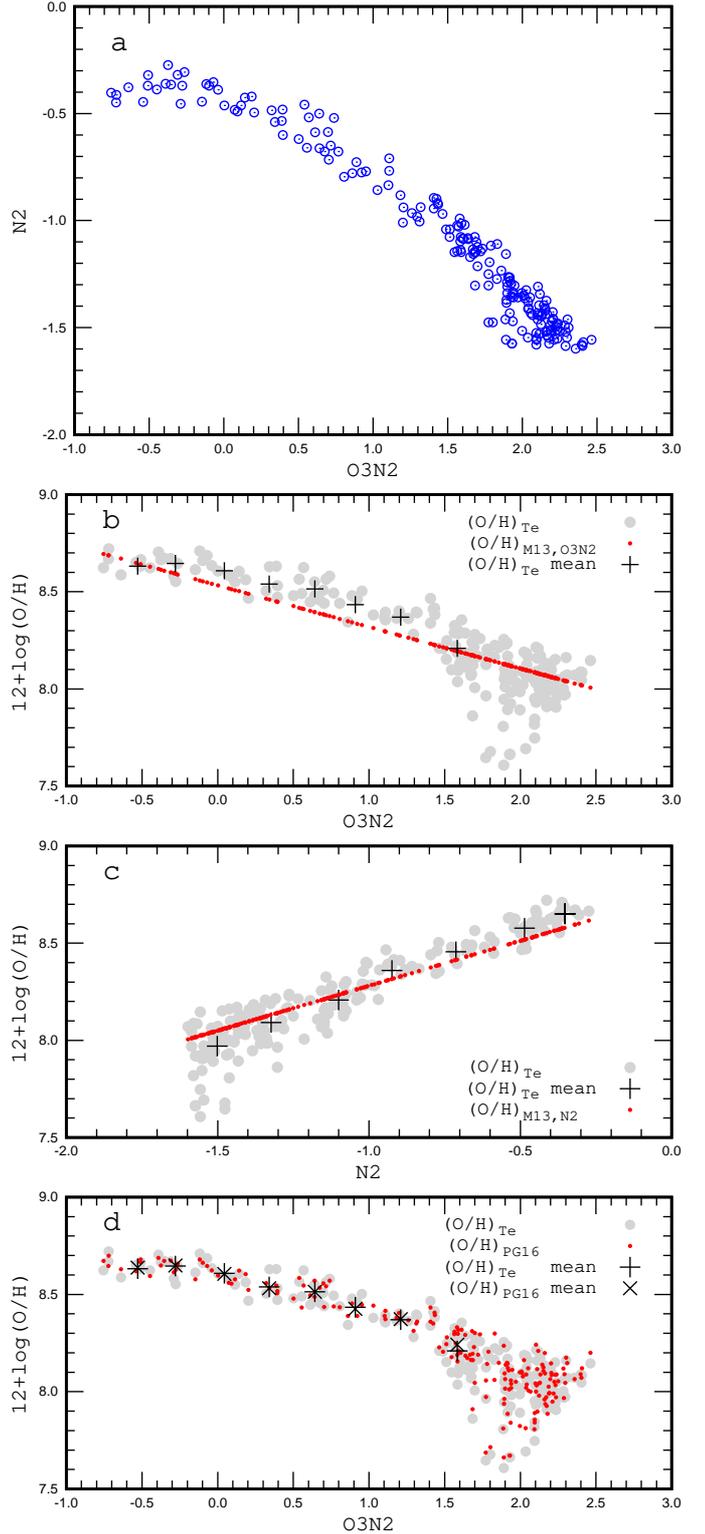}}
\end{center}
\caption{
Panel $a$ shows the N2 abundance index as a function of the O3N2 abundance index
for objects with $T_{e}$-based abundances (compilation from \citet{Pilyugin2016}).
Panel $b$ shows the (O/H)$_{T_{e}}$ abundances (grey points) and the (O/H)$_{\rm M13,O3N2}$
abundances (red points) for individual objects, and the mean (O/H)$_{T_{e}}$ abundances (plus signs) 
in bins with a width of 0.3 in O3N2 as a function of O3N2.
Panel $c$ shows the (O/H)$_{T_{e}}$ abundances (grey points) and the (O/H)$_{\rm M13,N2}$
abundances (red points) for individual objects, and the mean (O/H)$_{T_{e}}$ 
abundances (plus signs) in bins with a width of 0.2 in N2 as a function of N2.
Panel $d$ shows the (O/H)$_{T_{e}}$ abundances (grey points) and the (O/H)$_{\rm PG16}$
abundances (red points) for individual objects, the mean (O/H)$_{T_{e}}$ abundances (plus signs) and 
(O/H)$_{\rm PG16}$ abundances (crosses) with a bin width of 0.3 in O3N2 
as a function of O3N2.
}
\label{figure:o3n2-n2}
\end{figure}

A sample of objects with $T_{e}$-based abundances has been compiled
in \citet{Pilyugin2016}. The analysis of this sample can tell us something
about the validity of the distributions of the abundances determined through the
different calibrations and can clarify the origin of the disagreement between
the break radii in those distributions. Only objects for which  
the N2 calibration is applicable (N2 $> -1.6$) are included in the analysis. 
Panels $a$ in Fig.~\ref{figure:o3n2-n2} shows the N2 abundance index as a
function of the O3N2 index for a sample of objects with $T_{e}$-based
oxygen abundances.
A comparison of panels $a$ in Fig.~\ref{figure:o3n2-n2} with panels $d$ in Fig.~\ref{figure:example-rpg-rma} shows that the O3N2 -- N2 diagram
for objects from \citet{Pilyugin2016} also shows an appreciable deviation
from the linear relation similar to the O3N2 -- N2 diagrams for individual spaxels in
each galaxy.
Panel $b$ shows the (O/H)$_{T_{e}}$ abundances (grey points) and the (O/H)$_{\rm M13,O3N2}$
abundances (red points) for individual objects, and the mean (O/H)$_{T_{e}}$ abundances (plus signs) 
in bins with a width of 0.3 in O3N2 as a function of O3N2.
Panel $c$ shows the abundances in (O/H)$_{T_{e}}$ (grey points) and (O/H)$_{\rm M13,N2}$
(red points) for individual objects, and the mean (O/H)$_{T_{e}}$ 
abundances (plus signs) in bins with a width of 0.2 in N2 as a function of N2.
Panel $d$ shows the abundances in (O/H)$_{T_{e}}$ (grey points) and (O/H)$_{\rm PG16}$
(red points) for individual objects, the mean (O/H)$_{T_{e}}$ abundances (plus signs), and 
(O/H)$_{\rm PG16}$ abundances (crosses) for a bin width of 0.3 in O3N2 
as a function of O3N2.

The comparison between panels $b$ and $c$ of Fig.~\ref{figure:o3n2-n2} shows that
the N2 calibration can be applied to a larger number of objects than the O3N2 calibration.
Indeed, all the selected objects satisfy the criterion of the applicability of the N2
calibration, N2 $> -1.6$ (panel $c$ in Fig.~\ref{figure:o3n2-n2}). However, some
of those objects do not satisfy the criterion of the applicability of the
O3N2 calibration, O3N2 $<$ 1.7  (panel $b$ in Fig.~\ref{figure:o3n2-n2}).

Panel $b$ of Fig.~\ref{figure:o3n2-n2} shows that there is an 
appreciable difference between the (O/H)$_{\rm M13,O3N2}$ and (O/H)$_{T_{e}}$ abundances
and the difference changes with the O3N2 index in a nonlinear manner. 
Therefore the (O/H)$_{\rm M13,O3N2}$ -- $R_{g}$ and the (O/H)$_{T_{e}}$ -- $R_{g}$ diagrams
for the same galaxy can show different slopes of the gradients and 
breaks at different radii.
Inspection of panel $c$ of Fig.~\ref{figure:o3n2-n2} shows that there is also a 
considerable difference between the (O/H)$_{\rm M13,N2}$ and (O/H)$_{T_{e}}$ abundances. 
However, the difference changes with the N2 index in a linear manner. 
Therefore the (O/H)$_{\rm M13,N2}$ -- $R_{g}$ and (O/H)$_{T_{e}}$ -- $R_{g}$ diagrams 
for the same galaxy can show different slopes of the gradients but they should 
show breaks (if they exist) at similar radii.
Examination of panel $d$ of Fig.~\ref{figure:o3n2-n2} shows that the
(O/H)$_{\rm PG16}$ and (O/H)$_{T_{e}}$ abundances are in agreement over the whole interval 
of the O3N2 index. 
Therefore the (O/H)$_{\rm PG16}$ -- $R_{g}$ and (O/H)$_{T_{e}}$ -- $R_{g}$ diagrams
for the same galaxy can show similar slopes of the gradients and 
exhibit breaks at the same (or at least quite similar) radii, 
i.e. the (O/H)$_{\rm PG16}$ -- $R_{g}$ diagram allows us to determine a 
reliable slope of the gradients and a reliable value of the break radius.

Thus, the break radius derived from the (O/H)$_{\rm M13,N2}$ -- $R_{g}$ diagram 
is more reliable than the break radius obtained from the (O/H)$_{\rm M13,O3N2}$ -- $R_{g}$ diagram. 
This explains why the agreement between the values of the break radii in the (O/H)$_{\rm M13,N2}$ -- $R_{g}$ 
and (O/H)$_{PG16}$ -- $R_{g}$ diagrams is better that 
the agreement between the values of the break radius in the (O/H)$_{\rm M13,O3N2}$ -- $R_{g}$ 
and  (O/H)$_{PG16}$ -- $R_{g}$ , or in the (O/H)$_{\rm M13,O3N2}$ -- $R_{g}$ and the 
(O/H)$_{\rm M13,N2}$ -- $R_{g}$, diagrams (see Table~\ref{rbrb}). 
\citet{Marino2016} also note that the N2 calibration provides a better match to the 
abundances obtained through the $T_{e}$ method.

\section{Surface brightness profiles}

\subsection{Bulge-disc decomposition}

\begin{figure}
\resizebox{1.00\hsize}{!}{\includegraphics[angle=000]{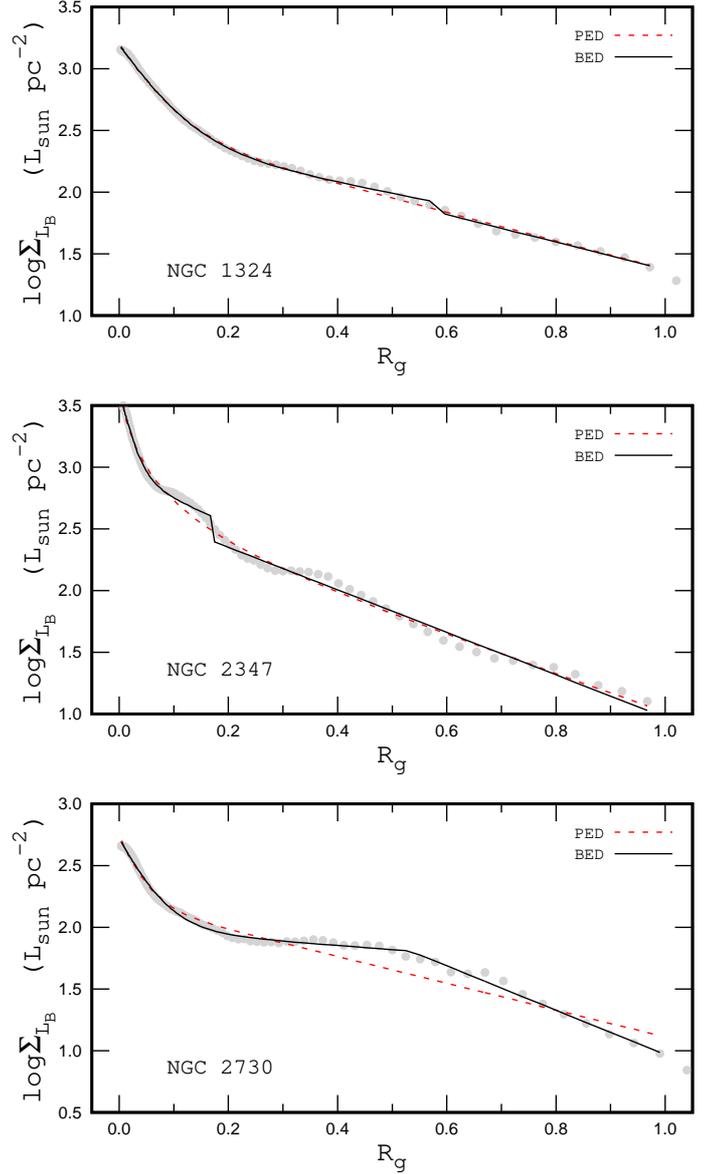}}
\caption{
Examples of the observed surface brightness profiles and fits. In each
panel, the grey points show the observed surface brightness profile.
The dashed (red) line indicates the pure exponential fit to those data. The
solid (black) line indicates the broken exponential fit.  The upper panel
shows the galaxy NGC~1324 with an observed surface brightness profile
close to a purely exponential disc.  The middle panels shows the galaxy
NGC~2347, which exhibits humps and troughs in the observed surface
brightness profile.  The lower panel shows the galaxy NGC~2730 with a
surface brightness profile with a clear break. 
}
\label{figure:galaxy-sb}
\end{figure}

Surface brightness measurements in solar units were used for the
bulge-disc decomposition. The magnitude of the Sun in the $B$ band of
the Vega photometric system, $B_{\sun} = 5.45$, was taken from
\citet{Blanton2007}. 

The stellar surface brightness distribution within a galaxy was 
fitted by an exponential profile for the disc and by a general
S\'{e}rsic profile for the bulge.  The total surface brightness
distribution was fitted with the expression 
\begin{eqnarray}
       \begin{array}{lll}
\Sigma_L(r) & = & (\Sigma_L)_{e}\exp \{-b_{n}[(r/r_{e})^{1/n} - 1]\} \\
            & + &  (\Sigma_{L})_{0}\exp(-r/h) ,                         \\
     \end{array}
\label{equation:decomp}
\end{eqnarray}
where $(\Sigma_L)_{e}$ is the surface brightness at the effective
radius $r_e$, i.e. the radius that encloses 50\% of the bulge light,
$(\Sigma_{L})_{0}$ is the central disc surface brightness, and $h$ the
radial scale length. The factor $b_n$ is a function of the shape
parameter $n$.  This factor can be estimated as $b_{n}  \approx 1.9992n  -
0.3271$ for $1 < n < 10$ \citep{Graham2001}.  The fit via
Eq.~(\ref{equation:decomp}) is referred to below as a purely
exponential disc (PED) approximation.  
It is known that the surface brightnesses in some galaxies
are flat or even increase out to a region of slope change where they
tend to fall off. Such surface brightness profiles can be formally
fitted by an exponential disc with a bulge-like component of negative
brightness \citep{Pilyugin2015}.

The parameters $(\Sigma_L)_{e}$, $r_e$, $n$, $(\Sigma_L)_{0}$, and
$h$ are obtained through the best fit to the observed radial surface
brightness profile, i.e. we derive a set of parameters in
Eq.~(\ref{equation:decomp}) that minimizes the deviation
$\sigma_{PED}$ of 
\begin{equation}
\sigma = \sqrt{ [\sum\limits_{j=1}^n (L(r_{j})^{cal}/L(r_{j})^{obs} - 1)^2]/n}  .
\label{equation:sigma}
\end{equation}
Here $L(r_{j})^{cal}$ is the surface brightness at the radius $r_{j}$
computed through Eq.~(\ref{equation:decomp}) and $L(r_{j})^{obs}$
is the measured surface brightness at that radius. 

The stellar surface brightness distribution within a galaxy was 
also fitted with the broken exponential  
\begin{eqnarray}
       \begin{array}{lll}
\Sigma_L(r) & = & (\Sigma_L)_{e}\exp \{-b_{n}[(r/r_{e})^{1/n} - 1]\} \\
            & + & (\Sigma_{L})_{0,inner}\exp(-r/h_{inner}) \;\;\;\; if \;\;\; r < R^{*} ,                        \\
            & = & (\Sigma_L)_{e}\exp \{-b_{n}[(r/r_{e})^{1/n} - 1]\} \\
            & + & (\Sigma_{L})_{0,outer}\exp(-r/h_{outer}) \;\;\;\; if \;\;\; r > R^{*}  .                        \\
     \end{array}
\label{equation:decomp2}
\end{eqnarray}
Here $R^{*}$ is the break radius, i.e. the radius at which the
exponent changes.  The fit via Eq.~(\ref{equation:decomp2}) is
referred to below as the broken exponential disc (BED)
approximation.  Again, eight parameters $(\Sigma_{L})_{e}$, $r_{e}$,
$n$, $(\Sigma_{L})_{0,inner}$, $h_{inner}$, $(\Sigma_{L})_{0,outer}$,
$h_{outer}$, and $R^{*}$ in the broken exponential disc were
determined through the best fit to the observed surface brightness
profile, i.e. we again require that the deviation $\sigma_{BED}$
given by Eq.~(\ref{equation:sigma}) is minimized.

\subsection{Breaks in the surface brightness profiles}

\begin{figure}
\resizebox{1.00\hsize}{!}{\includegraphics[angle=000]{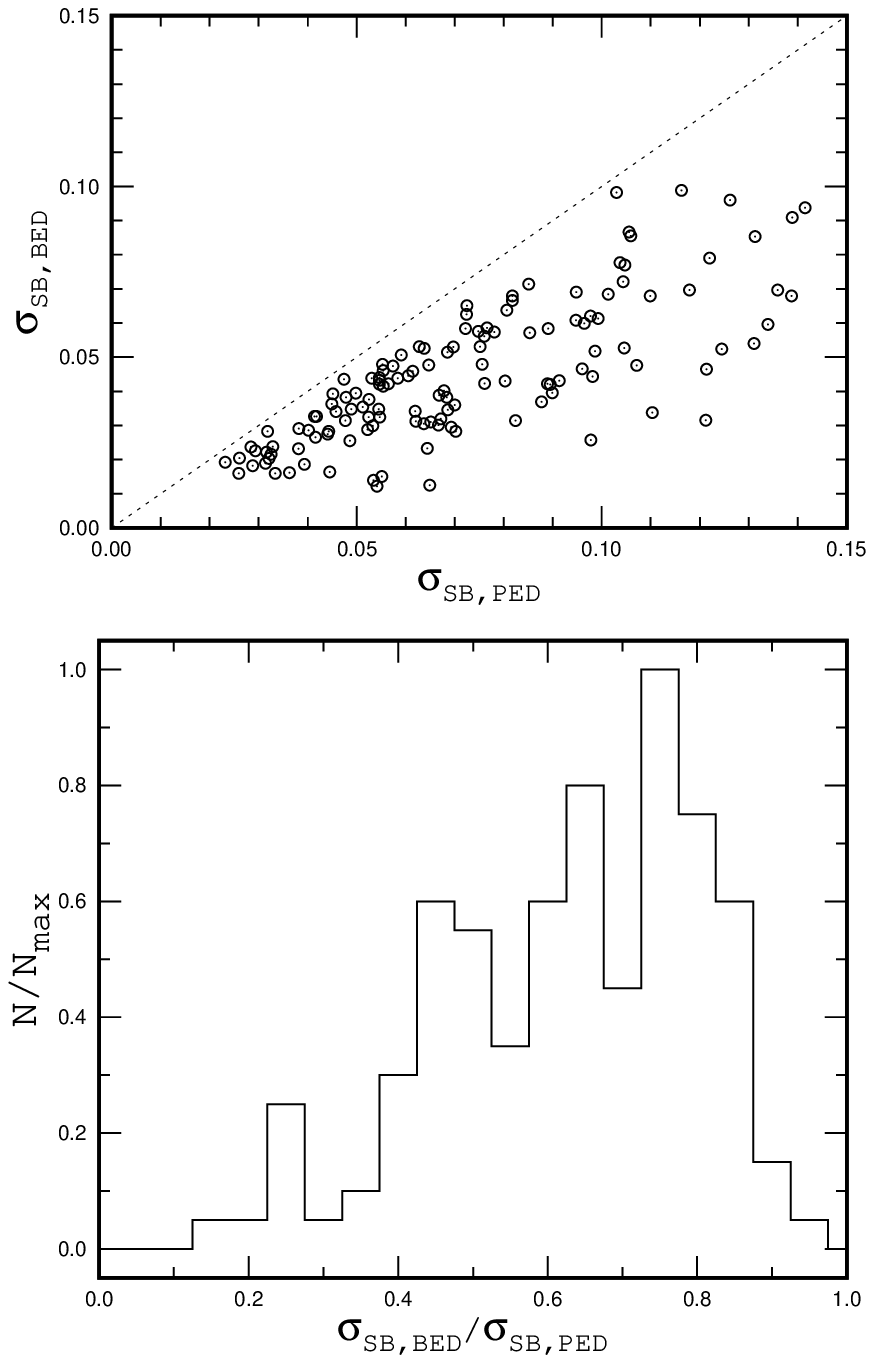}}
\caption{
In the upper panel the scatter of the surface brightness
$\sigma_{SB,BED}$  around the broken exponential fit vs. the
scatter $\sigma_{SB,PED}$ around the pure exponential fit is plotted.
The dashed line indicates equal values.
The lower panel shows the
normalized histogram of the scatter ratios
$\sigma_{SB,BED}$/$\sigma_{SB,PED}$ for our sample of galaxies. 
}
\label{figure:ssbp-ssbb}
\end{figure}

\begin{figure}
\resizebox{1.00\hsize}{!}{\includegraphics[angle=000]{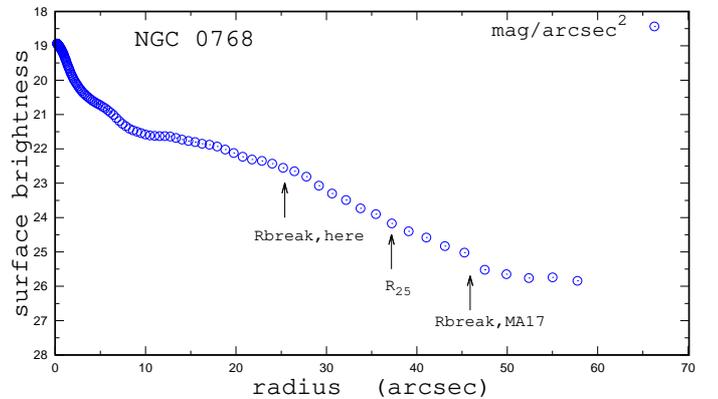}}
\caption{
Observed surface brightness profile of the galaxy NGC~0768 
in the $g$ SDSS photometric band (points).
The optical radius $R_{25}$ and the breaks in the surface brightness 
distribution obtained here and by \citet{MendezAbreu2017} are 
shown with the arrows. 
}
\label{figure:breaksMA}
\end{figure}

Fig.~\ref{figure:galaxy-sb} shows examples of the surface brightness
profiles of our sample of galaxies. The grey points are the observed
surface brightness profiles. The dashed (red) line is the purely
exponential fit to those data, the solid (black) line is the broken
exponential fit.  The upper panel of Fig.~\ref{figure:ssbp-ssbb} shows
the scatter of the surface brightnesses $\sigma_{SB,BED}$ around the
broken exponential disc profile as a function of scatter
$\sigma_{SB,PED}$ around the pure exponential disc profile.  The
circles stand for individual galaxies. The dashed line indicates equal
values. 

The upper panel of Fig.~\ref{figure:galaxy-sb} shows the measured
surface brightness profile of the galaxy NGC~1324, which is an example
of a galaxy with a surface brightness profile close to a pure
exponential disc.  The pure and broken exponential disc fits are
close to each other.  The scatter in the surface brightnesses around
the pure and broken exponential disc profiles are
rather small for those galaxies.  These galaxies are located near the
line of equal values in the $\sigma_{SB,BED}$ -- $\sigma_{SB,PED}$
diagram, Fig.~\ref{figure:ssbp-ssbb}, at low values of 
$\sigma_{SB,BED}$ and $\sigma_{SB,PED}$.  The scatter ratio
$\sigma_{SB,BED}$/$\sigma_{SB,PED}$ for those galaxies is close to 1.

The middle panel of Fig.~\ref{figure:galaxy-sb} shows the measured
surface brightness profile of the galaxy NGC~2347, which is an example
of a galaxy with humps and troughs in the observed surface brightness
profile. This may be a result of spiral arms or other structures.  The
scatter in the surface brightnesses both around the pure and broken exponential disc profiles is large for those galaxies but
the deviation of the scatter ratio $\sigma_{SB,BED}$/$\sigma_{SB,PED}$
from 1 is not very large. These galaxies
are located near the line of equal
values in the $\sigma_{SB,BED}$ -- $\sigma_{SB,PED}$ diagram,
Fig.~\ref{figure:ssbp-ssbb}, at large values of $\sigma_{SB,BED}$
and $\sigma_{SB,PED}$.

The lower panel of Fig.~\ref{figure:galaxy-sb} shows the measured
surface brightness profile of the galaxy NGC~2730, which is an example
of a galaxy where the surface brightness profile shows a prominent
break. The purely exponential disc is in significant disagreement with
the observed surface brightness profile; the value of the scatter
$\sigma_{SB,PED}$ is high. In contrast, the broken exponential disc
adequately reproduces the observed surface brightness profile, and the
value of the scatter $\sigma_{SB,BED}$ is low.  The scatter ratio
$\sigma_{SB,BED}$/$\sigma_{SB,PED}$ for those galaxies is
significantly less than 1.  These galaxies are located much below the
line of equal values on the $\sigma_{SB,BED}$ -- $\sigma_{SB,PED}$
diagram, Fig.~\ref{figure:ssbp-ssbb}, both at low and high values of
$\sigma_{SB,PED}$.

Thus, the scatter ratio $\sigma_{SB,BED}$/$\sigma_{SB,PED}$ for a
galaxy can be considered as an indicator of the presence of a break in
its surface brightness profile.  The lower panel of
Fig.~\ref{figure:ssbp-ssbb} shows the normalized histogram of the
scatter ratios $\sigma_{SB,BED}$/$\sigma_{SB,PED}$ for our sample of
galaxies. One can see that the $\sigma_{SB,BED}$/$\sigma_{SB,PED}$ ratio
in a number of galaxies is significantly lower than 1.   
The  $\sigma_{SB,BED}$/$\sigma_{SB,PED}$ ratio can serve as an indicator 
of the strength of the break in the surface brightness profile. 
Galaxies with $\sigma_{SB,BED}$/$\sigma_{SB,PED}$ less than 0.6 
in their surface brightness profiles are referred to as 
galaxies with a prominent break in their surface brightness profiles.
Certainly, the choice of the value of 0.6 is somewhat arbitrary. 

It should be emphasized that the surface brightness distributions
within the optical radii of galaxies are only considered and fitted
in our current study.  Recently, \citet{MendezAbreu2017} presented a 2D 
multi-component photometric decomposition of 404 CALIFA galaxies 
in the $g$, $r$, and $i$ SDSS bands. These authors considered the surface 
brightness distributions well beyond the optical radii of these galaxies. 
Fig.~\ref{figure:breaksMA} shows the observed surface brightness profile of 
the galaxy NGC~0768 in $g$ SDSS band using point symbols.
The optical radius $R_{25}$ and the breaks in the surface brightness 
distribution obtained here and by \citet{MendezAbreu2017} are 
shown with arrows.
Fig.~\ref{figure:breaksMA} illustrates that breaks at different 
radii can be found if the surface brightness profile is considered 
and fitted to the different limits of the galactocentric distances.

\subsection{Relation of breaks in surface brightness profiles and abundance gradients}

\begin{figure}
\resizebox{1.00\hsize}{!}{\includegraphics[angle=000]{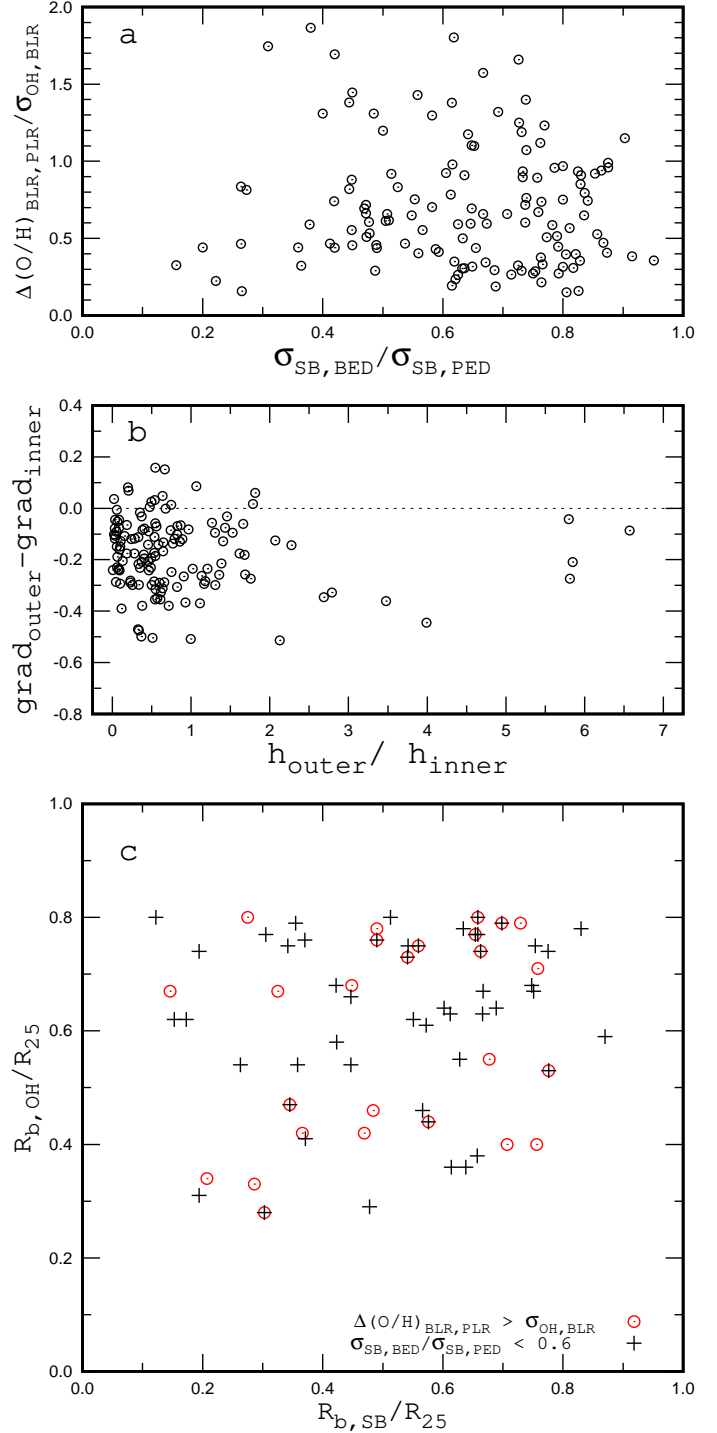}}
\caption{
Panel $a$ shows the difference between the broken and linear relation  
$\Delta$(O/H)$_{BLR,PLR}$/$\sigma_{OH,BLR}$ (indicator of the strength of 
break in the O/H distribution) as a function of $\sigma_{SB,BED}$/$\sigma_{SB,PED}$ 
(indicator of the strength of the break of the surface brightness profile).
Panel $b$ shows of the difference between abundance gradients on different 
sides of the break in the abundance distribution as a function of the ratio 
of the disc scale lengths on different sides of the break in the surface brightness 
profile.
Panel $c$ shows the break radius of the radial oxygen abundance gradient $R_{b,OH}$
vs. the break radius of the surface brightness profile $R_{b,SB}$.
Galaxies with a $\Delta$(O/H)$_{BLR,PLR}$/$\sigma_{OH,BLR}$ ratio larger than
1 (with a clear break in the radial abundance gradient) are plotted
with circles.  
Galaxies with a $\sigma_{SB,BED}$/$\sigma_{SB,PED}$ ratio less than
0.6 (with a prominent break in the surface brightness profile) are indicated
by plus signs.  
The break radii are normalized to the isophotal radii
$R_{25}$ of the galaxies.  
}
\label{figure:rb-rb}
\end{figure}

\begin{figure*}
\resizebox{0.90\hsize}{!}{\includegraphics[angle=000]{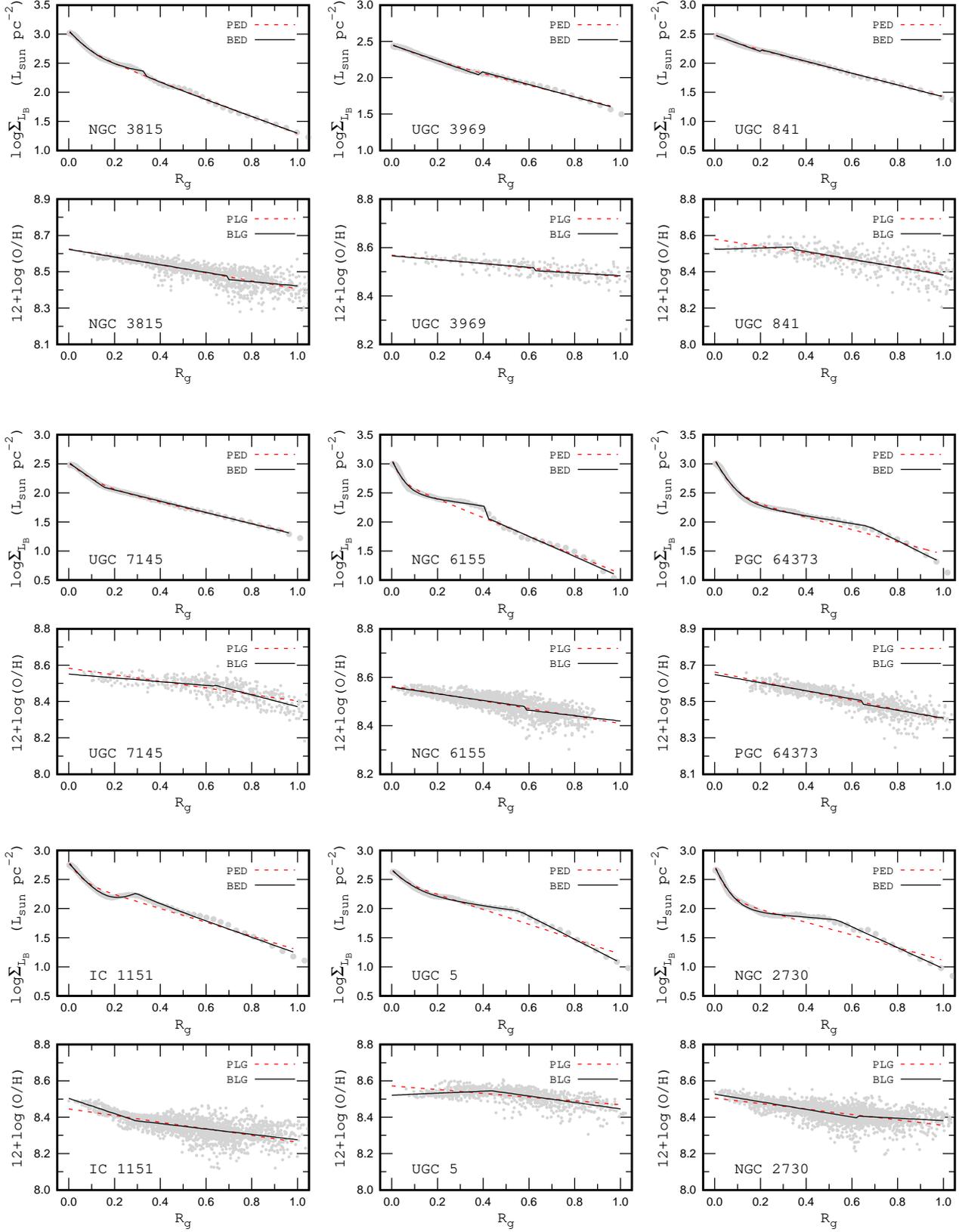}}
\caption{
Comparison between the shapes of surface brightness profile and
abundance gradient in galactic discs.  The grey points show the
observed surface brightnesses (abundances). The dashed (red) line is
the purely exponential (linear) fit to those data, the solid (black)
line is the broken exponential (linear) fit.  NGC~3815
and UGC~3969 are examples of galaxies where the surface brightness
profile is well fitted by a purely exponential disc and the abundance
gradient is well described by a purely linear relation.  
UGC~841 and UGC~7145 are examples of galaxies where the surface
brightness profile is well fitted by a purely exponential disc but the
abundance gradient shows a prominent break.  NGC~6155
and PGC~64373 are examples of galaxies where the surface brightness
profile shows an appreciable break while the abundance gradient is
well described by a purely linear relation.  IC~1151,
UGC~5, and NGC~2730 are examples of galaxies where both the surface
brightness profile and the abundance gradient display a prominent
break.
}
\label{figure:galaxy-sb-oh}
\end{figure*}

The break in both the surface brightness profile and abundance
distribution in the disc of a galaxy is specified by three caracteristics:
the strength (significance) of the break; the type of the break, i.e. the character of the change in the slope of the distribution when passing the break;
and by the break radius.
If the breaks in the surface brightness profile and abundance
gradient in the disc of a galaxy are related then one may expect that
the shapes of the surface brightness profiles and abundance gradients in the galactic discs should be similar in the
sense that galaxies with a prominent break in the surface
brightness profile should also display a prominent break in the radial
abundance gradient of the same type, and the position of the break in the
surface brightness profile should coincide with the position of the
break in the radial abundance gradient.

Panel $a$ in Fig.~\ref{figure:rb-rb} shows the difference between the 
broken and linear relation  $\Delta$(O/H)$_{BLR,PLR}$/$\sigma_{OH,BLR}$ , which is an indicator of the strength of break in the radial 
abundance distribution, as a function of the $\sigma_{SB,BED}$/$\sigma_{SB,PED}$ ratio, 
which is an indicator of the break strength in the surface brightness 
profile). There is no correlation between the strength of the break in the radial 
abundance distribution and the strength of the break in the surface brightness 
profile. A prominent break in the surface brightness profile 
(low value of the $\sigma_{SB,BED}$/$\sigma_{SB,PED}$ ratio) can be accompanied 
by a break of different strengths in the radial abundance distribution,  
and the weak break in the radial abundance distribution (low value of the  
$\Delta$(O/H)$_{BLR,PLR}$/$\sigma_{OH,BLR}$) can take place in galaxies with 
both strong and weak breaks in the surface brightness profiles.

Panel $b$ in Fig.~\ref{figure:rb-rb} shows the difference between 
abundance gradients on different sides of the break in the abundance 
distribution as a function of the ratio of the disc scale lengths on 
different sides of the break in surface brightness profile. 
The value of the $grad_{outer}$ -- $grad_{inner}$ specifies the type 
of the break in the radial abundance profile. 
The positive value of the difference $grad_{outer}$ -- $grad_{inner}$ 
shows that the gradient becomes flatter in going through the 
break point (up-bending radial abundance profile or looking-down 
break). The negative value of the difference $grad_{outer}$ -- $grad_{inner}$ 
shows that the gradient becomes steeper in going through the 
break point (down-bending radial abundance profile or looking-up 
break). 
The ratio $h_{outer}$/$h_{inner}$ specifies the type 
of the break in the surface brightness profile. 
A ratio $h_{outer}$/$h_{inner}$ lower than 1 means that the surface 
brightness profile steepens in going through the 
break point (down-bending surface brightness profile or looking-up 
break). 
A ratio $h_{outer}$/$h_{inner}$ higher than 1 means that the surface 
brightness profile flattens in going through the 
break point (up-bending surface brightness profile, or looking-down 
break). 
Inspection of panel $b$ in Fig.~\ref{figure:rb-rb} shows that
there is no correlation between the type of break in the radial 
abundance distribution and the type of break in the surface brightness 
profile. 
Indeed the down-bending surface brightness profile can be accompanied by 
a down-bending or by an up-bending radial abundance profile, 
and a down-bending radial abundance profile can be accompanied by both 
down-bending and up-bending surface brightness profiles.

Panel $c$ in Fig.~\ref{figure:rb-rb} shows the break radius of the radial oxygen
abundance gradient $R_{b,OH}$ versus the break radius of the surface
brightness profiles  $R_{b,SB}$. 
Galaxies with a $\Delta$(O/H)$_{BLR,PLR}$/$\sigma_{OH,BLR}$ ratio larger than
1 (with a clear break in the radial abundance gradient) are shown 
by circles.  
Galaxies with a $\sigma_{SB,BED}$/$\sigma_{SB,PED}$ ratio less than
0.6 (with a prominent break in the surface brightness profile) are indicated
by plus signs.  
The break radii are normalized to the isophotal radii
$R_{25}$ of the galaxies.  
Inspection of panel $c$ in Fig.~\ref{figure:rb-rb} shows that there is no
correlation between the break radius of the radial oxygen abundance
gradient $R_{b,OH}$ and the break radius of the surface brightness
profiles  $R_{b,SB}$.

Fig.~\ref{figure:galaxy-sb-oh} shows examples of galaxies with
different shapes of the surface brightness profiles and abundance
gradients.  The grey points denote the observed surface brightnesses
(abundances). The dashed (red) line is the purely exponential (linear)
fit to those data, the solid (black) line represents the broken
exponential (linear) fit.  The galaxies NGC~3815 and UGC~3969 are
examples of galaxies where the surface brightness profile of the disc
is close to a purely exponential disc, i.e. the profile is well
fitted by a purely exponential disc, and the abundance gradient is
well described by a purely linear relation.  The galaxies UGC~841 and
UGC~7145 are examples of galaxies where the surface brightness profile
is well fitted by a purely exponential disc but the abundance gradient
shows a prominent break.  Thus, the purely exponential profile of the
surface brightness may be accompanied by either a pure or a broken
linear profile of the radial oxygen abundance distribution.

The galaxies NGC~6155 and PGC~64373 in Fig.~\ref{figure:galaxy-sb-oh}
are examples of galaxies where the surface brightness profile shows an
appreciable break while the abundance gradient is well described by a
purely linear relation.  The galaxies IC~1151, UGC~5, and NGC~2730 are
examples of galaxies where both the surface brightness profile and abundance gradient display a prominent break.

Thus, the shape of the
surface brightness profile in the disc is not related to the shape
of the radial abundance gradient.  The broken exponential profile of
the surface brightness in the disc can be accompanied by either a
pure or broken linear profile of the radial oxygen abundance
distribution.  And vice versa, a purely exponential profile of the
surface brightnes can be accompanied by a pure or by a broken linear
profile of the radial oxygen abundance distribution.
A break in the surface brightness profile need not be
accompanied by a break in the radial abundance gradient.  The shape of
the surface brightness profile is thus independent of the shape of the
radial abundance gradient. 

\citet{Marino2016} have compared the abundance gradients in the inner
and outer disc parts divided by the break in the surface brightness profile
for the CALIFA galaxies. They do not derive the break radius in
the abundance gradient through a direct fit of the radial abundance distribution
by the broken relation. This prevents us from comparing our results with their results.

\section{Discussion}


\begin{figure}
\resizebox{1.00\hsize}{!}{\includegraphics[angle=000]{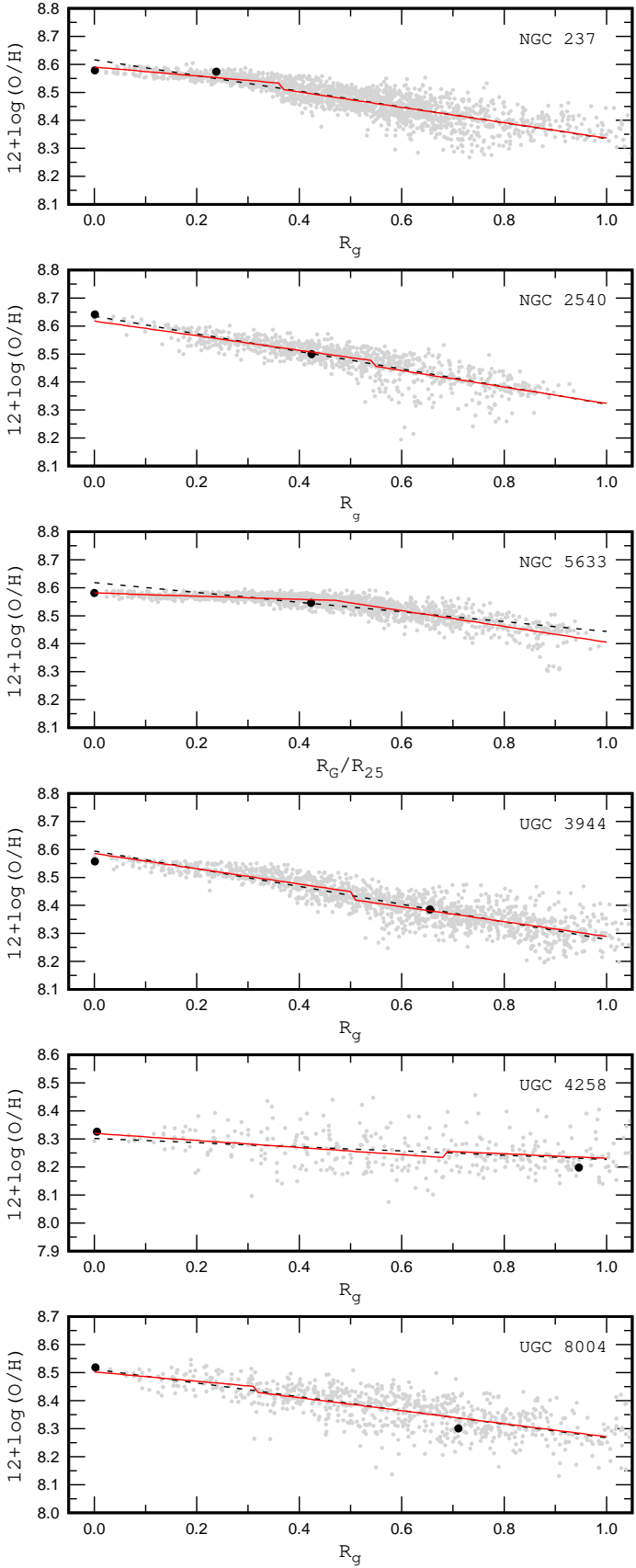}}
\caption{
Examples of the abundance distributions in the discs of galaxies with
available SDSS spectroscopy.  The grey points stand for abundances
in individual regions (spaxels) obtained from the CALIFA spectra.  The
dashed (black) line indicates a purely linear fit to those data; the solid
(red) line indicates a broken linear fit.  The dark (black) points denote the
oxygen abundances determined from the SDSS spectra. 
}
\label{figure:galaxy-oh-ohsdss}
\end{figure}

\begin{figure}
\resizebox{1.00\hsize}{!}{\includegraphics[angle=000]{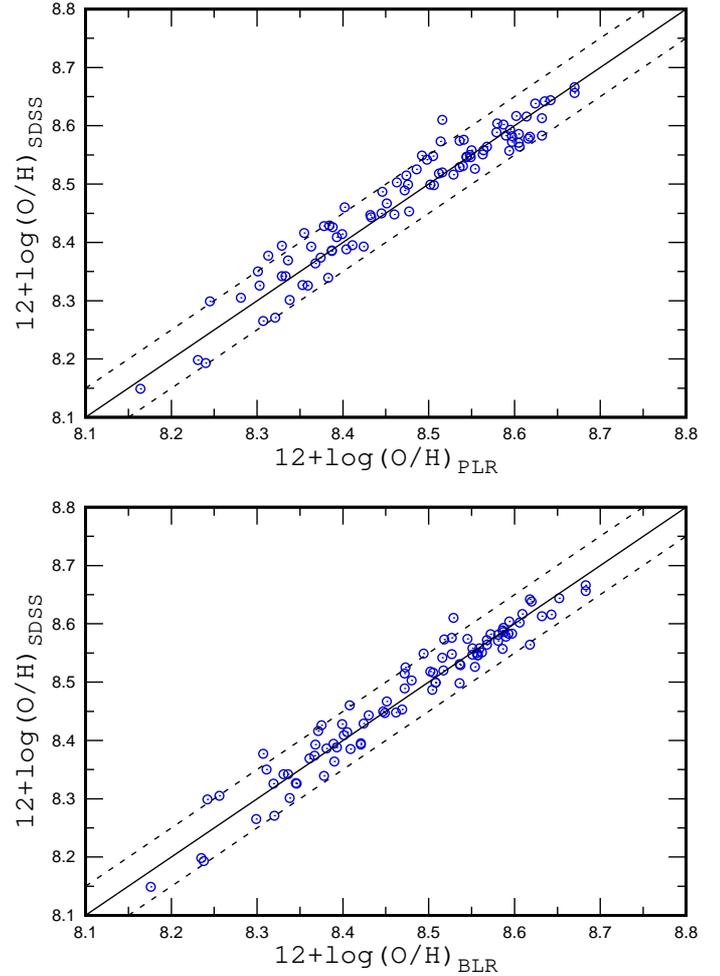}}
\caption{
Upper panel: The oxygen abundance (O/H)$_{SDSS}$ determined from the
SDSS spectrum vs.\ abundance (O/H)$_{PLR}$ given by the purely linear
O/H -- $R_{g}$ relation at this radius.  The solid line indicates
equal values. The dashed lines show the $\pm$0.05 dex deviations from
unity.  Lower panel: The same as the upper panel but for the
(O/H)$_{BLR}$ given by the broken linear O/H -- $R_{g}$ relation. 
}
\label{figure:ohsdss}
\end{figure}

Spectroscopic measurements of H\,{\sc ii} regions in a number of galaxies in
our sample are available in the SDSS database.  An SDSS spectrum of one 
region, usually near the centre of the galaxy, is available for around 
50 galaxies of our sample, and SDSS measurements of two regions at 
different galactocentric distances are available in 20 galaxies.  
This provides an additional possibility to check the validity of the oxygen 
abundances and radial abundance gradients estimated by us based on CALIFA 
survey spectroscopy.

The measurements of the emission-line fluxes (H$\beta$, [O\,{\sc
iii}]$\lambda$5007, H$\alpha$, [N\,{\sc ii}]$\lambda$6584, [S\,{\sc
ii}]$\lambda$6717, and [S\,{\sc ii}]$\lambda$6731) in the SDSS spectra
and the fiber coordinates were extracted from the SDSS DR12 data
\citep{Alam2015} reported in the {\sc ned}.  The demarcation line of
\citet{Kauffmann2003} between H\,{\sc ii} region-like and AGN-like
objects in the BPT classification diagram of \citet{Baldwin1981} was
adopted to select the H\,{\sc ii} region-like objects.  The
dereddening and abundance determinations were carried in the same
way as in the case of the CALIFA spectra. The fractional
galactocentric distances (normalized to the isophotal radius) of the
regions with SDSS spectra were estimated with the geometrical
parameters of galaxies obtained here. 

Fig.~\ref{figure:galaxy-oh-ohsdss} shows examples of the abundance
distributions in the discs of galaxies with available SDSS spectra.
The grey points denote the oxygen abundances in individual regions
(spaxels) determined using the CALIFA survey spectra. The dashed
(black) line indicates the purely linear fit to those data; the solid (red)
line indicates the broken linear fit.  The dark (black) points stand for
abundances in individual regions (fibers) estimated from the SDSS
spectra.  Fig.~\ref{figure:galaxy-oh-ohsdss} shows that the abundances
derived from the SDSS spectra are in agreement with the abundances
obtained from the CALIFA spectra.  All the regions with SDSS-based
oxygen abundances are located within the bands outlined by the regions
with CALIFA-based abundances. 

A quantitative comparison between the SDSS-based and CALIFA-based
abundances can be performed in the following manner.  Using the
galactocentric distance of the SDSS fibers, we determined the
corresponding abundance (O/H)$_{PLR}$  given by the purely linear O/H
-- $R_{g}$ relation at this galactocentric distance and the
abundance (O/H)$_{BLR}$ given by the broken linear O/H -- $R_{g}$
relation for that galaxy.  The upper panel of Fig.~\ref{figure:ohsdss}
shows the comparison between the (O/H)$_{SDSS}$ and (O/H)$_{PLR}$
abundances. The circles stand for the abundances in the individual
regions.  The solid line indicates equal values. The dashed lines
show the $\pm 0.05$ dex deviations from unity.  The upper panel of
Fig.~\ref{figure:ohsdss} shows that the SDSS-based abundances are
close to the purely linear relation O/H -- $R_{g}$ obtained from the
CALIFA-based abundances.  The deviations of the bulk of the SDSS-based
abundances from the CALIFA-based gradients are within 0.05 dex.  The
lower panel of Fig.~\ref{figure:ohsdss} shows the comparison between
the (O/H)$_{SDSS}$ and (O/H)$_{BLR}$ abundances.  This panel shows
that the SDSS-based abundances are also close to the broken linear
relation O/H -- $R_{g}$ determined using the CALIFA-based abundances.
Again, the deviations of the bulk of the SDSS-based abundances from
the CALIFA-based gradients are within 0.05 dex. 

Thus the oxygen abundances estimated from the SDSS spectra agree with
the abundances obtained from the CALIFA spectra within $\sim 0.05$ dex.
This is strong evidence supporting that the use of the spectra of
individual spaxels for abundance determinations is justified, our
measurements of the line fluxes are reliable, and the criterion
$\epsilon \geq 3$ for strong lines to select the spaxels for analysis 
does not introduce any bias in the determined abundance gradients.

It should be emphasized that in our current study abundance
gradients within the optical radii of galaxies are considered.
Spectroscopy of H\,{\sc ii} regions in the extended discs of several
spiral galaxies were carried out by
\citet{Bresolin2009,Goddard2011,Bresolin2012,Patterson2012}.  The
radial oxygen abundance gradients in those galaxies were estimated out
to around 2.5 times the optical isophotal radius. These authors found
that the slope of the radial oxygen abundance gradients changes at the
optical radii of their target galaxies, such that beyond the optical
radius of the disc the abundance gradients become flatter. In
addition, there appears to be an abundance discontinuity close to this
transition \citep{Bresolin2009,Goddard2011}.  The change in the
gradient slope is more distinct in the radial distribution of nitrogen
than in that of oxygen abundances \citep{Pilyugin2012}. 
\citet{SanchezMenguiano2016} found that most of the CALIFA galaxies 
with reliable oxygen abundance values beyond $\sim$2 effective radii 
present a flattening of the abundance gradient in these outer regions. 
They suggest that such flattening seems to be a universal property of 
spiral galaxies.

It is known that there is a correlation between the local oxygen
abundance and stellar surface brightness of a galactic disc
\citep[e.g.][]{Webster1983,Edmunds1984,Ryder1995,Moran2012,RosalesOrtegaetal2012,Pilyugin2014b}.
If this correlation is local, i.e. there is a point-to-point
correlation, then one may expect that the break in the surface
brightness profile should be accompanied by a break in the radial
abundance gradient.  We found in our current study that the shapes of
the abundance gradient and surface brightness profile may be
different in a given galaxy in the sense that a broken exponential
surface brightness profile of a disc may be accompanied by either a
pure or broken linear profile of the radial oxygen abundance
distribution.  Vice versa, a disc with a pure exponential profile of
the surface brightness may show either a pure or broken linear
profile of the radial oxygen abundance distribution.  We also found
that there is no correlation between the break radii of the abundance
gradient and surface brightness profiles.  This suggests that
there is no unique point-to-point correlation  
between local abundance and local surface brightness
in the discs of galaxies.

Another argument against a unique point-to-point correlation between
the local oxygen abundance and stellar surface brightness in
galaxies comes from the following general consideration.
By definition, the values of the stellar surface brightness at the
optical radius should be very close in all the galaxies. However, the
oxygen abundances at the optical radius vary strongly from galaxy to
galaxy. 
The oxygen abundances at the optical radius 12+log(O/H)$_{R_{25}}$ vary from
$\sim$8.1 to $\sim$8.6 for the galaxies of our sample. That is, the variation in
the oxygen abundances at the optical radii of different galaxies at a fixed
surface brightness is comparable to, or exceeds, the radial variation in
the oxygen abundance within a given galaxy; there is a variation of the surface
brightness by two to three orders of magnitude.
Thus, the comparison of the oxygen
abundances at the optical radii of different galaxies provides 
prominent evidence against a unique point-to-point correlation between
the local oxygen abundance and stellar surface brightness.
This supports our conclusion
that the correlation between the local oxygen abundance and stellar surface
brightness is caused by a relation between the global abundance
distribution and global surface brightness distribution rather than by
a local, point-to-point correlation.

\section{Conclusions}

We constructed maps of the oxygen abundance in the discs of 134 spiral
galaxies using the 2D spectroscopy of the DR3 of the CALIFA survey.
The radial abundance gradients within the optical isophotal radii were
determined.  To examine the existence of a break in the slope of the
gradient, the radial oxygen abundance distribution in each galaxy was
fitted by a purely linear relation O/H = $f(R_{g}),$ where the
commonly accepted logarithmic scale for the oxygen abundance was used,
and by a broken linear relation. 

Photometric maps of our target galaxies in the $B$ band were constructed using
publicly available photometric measurements in the SDSS $g$ and $r$
bands. We carried out bulge-disc decompositions of the obtained surface
brightness profiles.  The stellar surface brightness distribution
across the galaxies was fitted by an exponential profile for the discs
and by a general S\'{e}rsic profile for the bulges.  The surface
brightness distribution across the discs was fitted with a pure and a broken exponential.  

We found that the maximum absolute difference between the oxygen
abundances in a disc given by the broken and purely linear
relations is less than 0.05 dex for the majority of our galaxies
and exceeds the scatter in abundances for 26 out of 134 galaxies considered. 
The scatter in abundances around the broken relation is close to the
scatter around the purely linear relation; the difference is usually
within 5\%.  Our results suggest that a simple linear relation is
adequate to describe the radial oxygen abundance distribution in the
discs of spiral galaxies and can be used (at least as the first order
approximation) for many tasks.  

The breaks in the surface brightness profiles in some galaxies are
more prominent; the scatter around the broken exponent is lower by a
factor of two and more than that around the pure exponent.  The ratio of
the scatter around the broken and pure exponential fits can be
considered as an indicator of the presence of a break in the surface
brightness profile.

The shapes of the surface brightness profile of the disc and its
abundance gradient can differ in the sense that the broken
exponential profile of the surface brightness in the disc can be
accompanied by either a pure or a broken linear profile of the radial
oxygen abundance distribution, and vice versa, the pure exponential
profile of the surface brightness can be accompanied by either a pure or 
a broken linear profile of the radial oxygen abundance distribution.
We also found that there is no correlation between the break radii of
the abundance gradient and surface brightness profiles.  Those results
demonstrate that a break in the surface brightness profile need not be
accompanied by a break in the abundance gradient.

Those results also suggest that
there is no unique point-to-point correlation  
between the local abundance and local surface brightness
in the discs of galaxies.
A significant variation in the oxygen abundances at the
optical radii of different galaxies (at a fixed surface brightness)
confirms this conclusion.

\section*{Acknowledgements}

We are grateful to the referee for his/her constructive comments. \\
L.S.P., E.K.G., and I.A.Z.\  acknowledge support within the framework
of Sonderforschungsbereich (SFB 881) on ``The Milky Way System''
(especially subproject A5), which is funded by the German Research
Foundation (DFG). \\ 
L.S.P.\ and I.A.Z.\ thank the 
Astronomisches Rechen-Institut at Heidelberg University where part of
this investigation was carried out for its hospitality. \\
I.A.Z. acknowledges the support of the Volkswagen Foundation 
under the Trilateral Partnerships grant No.\ 90411. \\
This work was partly funded by a subsidy allocated to Kazan Federal 
University for the state assignment in the sphere of scientific 
activities (L.S.P.).  \\ 
This study uses data provided by the Calar Alto Legacy Integral Field Area 
(CALIFA) survey (http://califa.caha.es/). 
Based on observations collected at the Centro Astron\'omico Hispano
Alem\'an (CAHA) 
at Calar Alto, operated jointly by the Max-Planck-Institut f\"ur Astronomie and 
the Instituto de Astrof\'{i}sica de Andaluc\'{i}a (CSIC). \\ 
We acknowledge the usage of the HyperLeda database (http://leda.univ-lyon1.fr). \\
This research made
use of Montage, funded by the National Aeronautics and Space
Administration's Earth Science Technology Office, Computational
Technnologies Project, under Cooperative Agreement Number NCC5-626
between NASA and the California Institute of Technology. The code is
maintained by the NASA/IPAC Infrared Science Archive. \\ 
Funding for SDSS-III has been provided by the Alfred P.\ Sloan
Foundation, the Participating Institutions, the National Science
Foundation, and the U.S. Department of Energy Office of Science. The
SDSS-III web site is http://www.sdss3.org/.


\begin{thebibliography}{}

\bibitem [Alloin et al.(1979)]{Alloin1979} 
          Alloin D., Collin-Souffrin S., Joly M., Vigroux L., 1979, A\&A, 78, 200


\bibitem [Ahn et al.(2012)]{Ahn2012} 
          Ahn C.P., Alexandroff R., Allende Prieto C., et al., 2012, ApJS, 203, 21

\bibitem[Alam et al.(2015)]{Alam2015} Alam, S., Albareti, F.~D., Allende 
         Prieto, C., et al.\ 2015, ApJS, 219, 12 

\bibitem [Andrievsky et al.(2016)]{Andrievsky2016}
         Andrievsky S.M., Martin R.P.,  Kovtyukh V.V., Korotin S.A., L\'{e}pine J.R.D., 2016, MNRAS, 461, 4256 
        
\bibitem [Baldwin et al.(1981)]{Baldwin1981}
          Baldwin J.A., Phillips M.M., \& Terlevich R. 1981, PASP, 93, 5

\bibitem [Belfiore et al.(2017)]{Belfiore2017} 
          Belfiore F., Maiolino R., Tremonti C., et al., 2017, MNRAS, 469, 151 

\bibitem [Berg et al.(2015)]{Berg2015}
          Berg D.A., Skillman E.D., Croxall K.V., Pogge R.W., Moustakas J., Johnson-Groh M., 2015, ApJ, 806, 16

\bibitem [Blanton \& Roweis(2007)]{Blanton2007} 
          Blanton, M. R., \& Roweis, S. 2007, AJ, 133, 734 

\bibitem [Bresolin et al.(2009)]{Bresolin2009}  
          Bresolin F., Ryan-Weber E., Kennicutt R.C., Goddard Q.,  2009, ApJ, 695, 580
          
\bibitem [Bresolin et al.(2012)]{Bresolin2012}  
          Bresolin F.,  Kennicutt R.C., Ryan-Weber E., 2012, ApJ, 750, 122
          
\bibitem [Bresolin \& Kennicutt(2015)]{BresolinKennicutt2015}
          Bresolin F., Kennicutt R.C., 2015, MNRAS, 454, 3664

\bibitem [Bundy et al.(2015)]{Bundy2015}
          Bundy K., Bershady M.A., Law D.R., et al., 2015, ApJ, 798, 7 

\bibitem [Croxall et al.(2015)]{Croxall2015}
          Croxall K.V., Pogge R.W.,  Berg D.A., Skillman E.D., Moustakas J.,  2015, ApJ, 808, 42

\bibitem [Croxall et al.(2016)]{Croxall2016}
          Croxall K.V., Pogge R.W.,  Berg D.A., Skillman E.D., Moustakas J.,  2016, ApJ, 830, 4 

\bibitem [de Vaucouleurs(1959)]{Vaucouleurs1959}
          de Vaucouleurs, G., 1959, Handbuch der Physics, 53, 311

\bibitem [Dinerstein(1990)]{Dinerstein1990} 
         Dinerstein H.~L.\ 1990, in The Interstellar Medium in
         Galaxies, ed.\ H.~A ~Thronson Jr.\ \& J.~M.~Shull 
         (Astrophysics and Space Science Library, Vol.\ 161; Dordrecht: 
         Kluwer), 257


\bibitem [Edmunds \& Pagel(1984)]{Edmunds1984} 
          Edmunds M.G., Pagel B.E.J., 1984, MNRAS, 211, 507 

\bibitem [Freeman(1970)]{Freeman1970}  
          Freeman K.C., 1970, ApJ, 160, 811

\bibitem [Garc\'{i}a-Benito et al.(2015)]{GarciaBenito2015}
          Garc\'{i}a-Benito R., Zibetti S., S\'{a}nchez S.F., et al., 2015, A\&A, 576, A135

\bibitem [Goddard et al.(2011)]{Goddard2011}  
          Goddard Q.E., Bresolin F., Kennicutt R.C., Ryan-Weber E.V., Rosales-Ortega F.F., 2011, MNRAS, 412, 1246

\bibitem [Graham(2001)]{Graham2001} 
          Graham A.W., 2001, AJ, 121, 820 

\bibitem [Gusev et al.(2012)]{Gusev2012}  
          Gusev A.S., Pilyugin L.S., Sakhibov F., Dodonov S.N., Ezhkova O.V., Khramtsova M.S., 
          2012, MNRAS, 424, 1930

\bibitem [Ho et al.(2015)]{Ho2015}  
          Ho I.-T., Kudritzki R.-P., Kewley L.J., Zahid H.J., Dopita M.A., Bresolin F., Rupke D.S.N., 2015, MNRAS, 448, 2030 

\bibitem [Husemann et al.(2013)]{Husemann2013}
          Husemann B., Jahnke K., S\'{a}nchez S.F., et al., 2013, A\&A, 549, A87

\bibitem [Izotov et al.(1994)]{Izotov1994}
          Izotov Y.I., Thuan T.X., Lipovetsky V.A., 1994, ApJ, 435, 647

\bibitem [Kauffmann et al.(2003)]{Kauffmann2003}
          Kauffmann G., Heckman T.M., Tremonti C., et al. 2003, MNRAS, 346, 1055

\bibitem [Kennicutt \& Garnett(1996)]{KennicuttGarnett1996}
          Kennicutt R.C., Garnett D.R., 1996, ApJ, 456, 504

\bibitem [Kewley et al.(2001)]{Kewley2001}
          Kewley L.J., Dopita M.A., Sutherland R.S., Heisler C.A., Trevena J.  2001 ApJ, 556, 121

\bibitem [Makarov et al.(2014)]{Makarov2014}
          Makarov D., Prugniel P., Terekhova N., Courtois H., Vauglin I., 2014, A\&A, 570, A13 

\bibitem [Marino et al.(2013)]{Marino2013}
          Marino R.A., Rosales-Ortega F.F., S\'{a}nchez S.F., et al., 2013, A\&A, 559, A114

\bibitem [Marino et al.(2016)]{Marino2016}
          Marino R.A., Gil de Paz A., S\'{a}nchez S.F., et al., 2016, A\&A, 585, A47 

\bibitem [Martin \& Roy(1995)]{MartinRoy1995}
          Martin P., Roy J.-R., 1995, ApJ, 445, 161

\bibitem [Martin et al.(2015)]{Martin2015}
          Martin R.P., Andrievsky S.M., Kovtyukh V.V., Korotin S.A., Yegorova I.A., Saviane I., 2015, MNRAS, 449, 4071

\bibitem [M\'{e}ndez-Abreu et al.(2017)]{MendezAbreu2017}
          M\'{e}ndez-Abreu J., Ruiz-Lara T., S\'{a}nchez-Menguiano L., et al., 2017, A\&A, 598, A32 

\bibitem [Moran et al.(2012)]{Moran2012} 
          Moran S.M., Heckman T.M., Kauffmann G., et al., 2012, ApJ, 745, 66 
        
\bibitem [Moustakas et al.(2010)]{Moustakas2010} 
          Moustakas J., Kennicutt R.C., Tremonti C.A., Dale D.A., 
          Smith J.-D.T., Calzetti D., 2010, ApJS, 190, 233 


\bibitem [Pagel et al.(1979)]{Pagel1979} 
          Pagel B.E.J., Edmunds M.G., Blackwell D.E., Chun M.S., Smith G., 1979, MNRAS, 189, 95


\bibitem [Patterson et al.(2012)]{Patterson2012} 
          Patterson M.T., Walterbos R.A.M., Kennicutt R.C., Chiappini C., Thilker D.A., 2012, MNRAS, 422, 401 

\bibitem [Paturel et al.(2003)]{Paturel2003} 
          Paturel G., Petit C., Prugniel P., et al., 2003, A\&A, 412, 45 

\bibitem [Peng et al.(2010)]{Peng2010} 
          Peng C.Y., Ho L.C., Impey C.D., Rix H.-W., 2010, AJ, 139, 2097 

\bibitem [Pettini \& Pagel(2004)]{Pettini2004} 
          Pettini M., Pagel B.E.J., 2014, MNRAS, 348, L59 

\bibitem [Pilyugin (2001)]{Pilyugin2001} 
          Pilyugin L.S., 2001, A\&A, 373, 56  

\bibitem [Pilyugin (2003)]{Pilyugin2003} 
          Pilyugin L.S., 2003, A\&A, 397, 109 

\bibitem [Pilyugin et al.(2004)]{Pilyugin2004} 
          Pilyugin L.S., V\'{\i}lchez J.M., Contini T., 2004, A\&A, 425, 849 

\bibitem [Pilyugin et al.(2006)]{Pilyugin2006} 
          Pilyugin L.S., Thuan T.X., V\'{\i}lchez J.M., 2006, MNRAS, 367, 1139 

\bibitem [Pilyugin et al.(2007)]{Pilyugin2007} 
          Pilyugin L.S., Thuan T.X., V\'{\i}lchez J.M., 2007, MNRAS, 376, 353 

\bibitem [Pilyugin et al.(2012)]{Pilyugin2012} 
          Pilyugin L.S., Grebel E.K., Mattsson L., 2012, MNRAS, 424, 2316

\bibitem [Pilyugin et al.(2014a)]{Pilyugin2014a} 
          Pilyugin L.S., Grebel E.K., Kniazev A.Y., 2014a, AJ, 147, 131 

\bibitem [Pilyugin et al.(2014b)]{Pilyugin2014b} 
          Pilyugin L.S., Grebel E.K., Zinchenko I.A., Kniazev A.Y., 2014b, AJ, 148, 134 

\bibitem [Pilyugin et al.(2015)]{Pilyugin2015} 
          Pilyugin L.S., Grebel E.K., Zinchenko I.A., 2015, MNRAS, 450, 3254
          
\bibitem [Pilyugin \& Grebel(2016)]{Pilyugin2016} 
          Pilyugin L.S., Grebel E.K., 2016, MNRAS, 457, 3678 

\bibitem [Pohlen \& Trujillo(2006)]{Pohlen2006} 
          Pohlen M., Trujillo I., 2006, A\&A, 454, 759 

\bibitem [Rosales-Ortega et al.(2012)]{RosalesOrtegaetal2012} 
          Rosales-Ortega F.F., S\'{a}nchez S.F., Iglesias-P\'{a}ramo J.,  et al. 2012, ApJ, 756, L31 

\bibitem [Ryder(1995)]{Ryder1995} 
          Ryder S.D., 1995, ApJ, 444, 610 

\bibitem [S\'{a}nchez et al.(2012)]{Sanchez2012} 
          S\'{a}nchez S.F., Kennicutt R.C., Gil de Paz A., et al.,  2012, A\&A, 538, A8

\bibitem [S\'{a}nchez et al.(2014)]{Sanchez2014}  
          S\'{a}nchez S.F., Rosales-Ortega F.F., Iglesias-P\'{a}ramo J., et al.  2014, A\&A, 563, 49 

\bibitem [S\'{a}nchez-Menguiano et al.(2016)]{SanchezMenguiano2016}  
          S\'{a}nchez-Menguiano L.,  S\'{a}nchez S.F., P\'{e}rez I., et al.  2016, A\&A, 587, A70 

\bibitem [Scarano et al.(2011)]{Scarano2011} 
          Scarano S., L\'{e}pine J.R.D., Marcon-Uchida M.M., 2011, MNRAS, 412, 1741

\bibitem [Schlafly \& Finkbeiner(2011)]{Schlafly2011} 
          Schlafly E. F., \& Finkbeiner D. P. 2011, ApJ, 737, 103 
        
\bibitem [Searle(1971)]{Searle1971} 
          Searle L. 1971, ApJ, 168, 327

\bibitem [Smith(1975)]{Smith1975}   
          Smith H.E. 1975, ApJ, 199, 591

\bibitem [Storey \&  Zeippen(2000)]{Storey2000}
          Storey P.J., Zeippen C.J., 2000, MNRAS, 312, 813

\bibitem[Thuan al.(2010)]{Thuan2010} 
          Thuan T.X., Pilyugin L.S., Zinchenko I.A., 2010, ApJ, 712, 1029 

\bibitem [van der Kruit(1979)]{vanderKruit1979}
          van der Kruit P.C., 1979, A\&AS, 38, 15 

\bibitem [van Zee et al.(1998)]{vanZee1998}
          van Zee L., Salzer J.J., Haynes M.P., O'Donoghue A.A., Balonek T.J., 1998, AJ, 116, 2805
         
\bibitem [Vila-Costas \& Edmunds(1992)]{VilaCostas1992} 
          Vila-Costas M.B., Edmunds M.G. 1992, MNRAS, 259, 121

\bibitem [Webster \& Smith(1983)]{Webster1983} 
          Webster B.L., Smith M.G., 1983, MNRAS, 204, 743 

\bibitem [York et al.(2000)]{York2000}
          York D.G., Adelman J., Anderson J.E., et al., 2000, AJ, 120, 1579

\bibitem [Zahid \& Bresolin(2011)]{Zahid2011} 
          Zahid H.J., Bresolin F., 2011, AJ, 141, 192

\bibitem [Zaritsky(1992)]{Zaritsky1992} 
          Zaritsky D., 1992, ApJ, 390, L73

\bibitem [Zaritsky et al.(1994)]{Zaritsky1994} 
          Zaritsky D., Kennicutt R.C., Huchra J.P., 1994, ApJ, 420, 87 

\bibitem [Zinchenko et al.(2015)]{Zinchenko2015} 
          Zinchenko I.A., Kniazev A.Y., Grebel E.K., Pilyugin L.S., 2015, A\&A, 582, A35 

\bibitem [Zinchenko et al.(2016)]{Zinchenko2016}
          Zinchenko I.A., Pilyugin L.S., Grebel E.K., S\'{a}nchez S.F., V\'{i}lchez J.M., 2016, MNRAS, 462, 2715 

 
 \end{thebibliography}
\end{document}